\documentclass{aastex63}
\usepackage{textcomp}
\usepackage{comment}
\usepackage{amsmath}
\usepackage{float}
\def\lea{\mathrel{<\kern-1.0em\lower0.9ex\hbox{$\sim$}}}
\def\gea{\mathrel{>\kern-1.0em\lower0.9ex\hbox{$\sim$}}}

\graphicspath{{./}{./Figures/}{figures/}}
\received{}
\revised{}
\accepted{}

\newcommand{\UTA}{\affiliation{Instituto de Alta Investigación, Universidad de Tarapacá, Casilla 7D, Arica, Chile}}
\newcommand{\UWyoming}{\affiliation{Department of Physics and Astronomy, University of Wyoming, Laramie, WY 82071, USA}}
\newcommand{\STScI}{\affiliation{Space Telescope Science Institute, 3700 San Martin Drive, Baltimore, MD 21218, USA}}
\newcommand{\UToledo}{\affiliation{Ritter Astrophysical Research Center, University of Toledo, Toledo, OH 43606, USA}}
\newcommand{\JHU}{\affiliation{Department of Physics and Astronomy, The Johns Hopkins University, Baltimore, MD 21218 USA}}
\newcommand{\Caltech}{\affiliation{TAPIR, California Institute of Technology, Pasadena, CA 91125 USA}}
\newcommand{\OSU}{\affiliation{Department of Astronomy, The Ohio State University, 140 West 18th Ave., Columbus, OH 43210, USA}}
\newcommand{\MPIA}{\affiliation{Max Planck Institut f\"ur Astronomie, K\"onigstuhl 17, 69117 Heidelberg, Germany}}
\newcommand{\UAlberta}{\affiliation{Department of Physics, University of Alberta, Edmonton, AB T6G 2E1, Canada}}
\newcommand{\Bonn}{\affiliation{Argelander-Institut für Astronomie, Universität Bonn, Auf dem Hügel 71, 53121, Bonn, Germany}}
\newcommand{\AthreeD}{\affiliation{ARC Centre of Excellence for All Sky Astrophysics in 3 Dimensions (ASTRO 3D), Australia}}
\newcommand{\ITA}{\affiliation{Institut f\"{u}r Theoretische Astrophysik, Zentrum f\"{u}r Astronomie der Universit\"{a}t Heidelberg,\\ Albert-Ueberle-Strasse 2, 69120 Heidelberg, Germany}}
\newcommand{\IWR}{\affiliation{Universit\"{a}t Heidelberg, Interdisziplin\"{a}res Zentrum f\"{u}r Wissenschaftliches Rechnen, Im Neuenheimer Feld 205, D-69120 Heidelberg, Germany}}
\newcommand{\STScIESA}{\affiliation{AURA for the European Space Agency (ESA), Space Telescope Science Institute, 3700 San Martin Drive, Baltimore, MD 21218, USA}}
\newcommand{\OHHS}{\affiliation{Ottawa Hills High School, 4035 W. Central Ave, Ottawa Hills, OH, 43606}}
\newcommand{\UMICHD}{\affiliation{Department of Natural Sciences, University of Michigan-Dearborn, 4901 Evergreen Road, Dearborn, MI 48128, USA}}
\newcommand{\UIOWA}{\affiliation{Department of Physics and Astronomy, University of Iowa, Iowa City, Iowa 52242, USA}}
\newcommand{\CWRU}{\affiliation{Department of Astronomy, Case Western Reserve University, 10900 Euclid Avenue, Cleveland, OH 44106, USA}}

\shorttitle{PHANGS-HST: Globular Cluster Systems}
\shortauthors{Floyd et al.}

\begin{document}

\title{PHANGS-HST: Globular Cluster Systems in 17 Nearby Spiral Galaxies}

\author[0009-0008-2929-6665]{Matthew~Floyd}
\author[0000-0003-0085-4623]{Rupali~Chandar}
\UToledo
\author{Bradley~C.~Whitmore}
\STScI
\author[0000-0002-8528-7340]{David~A.~Thilker}
\JHU
\author[0000-0003-0946-6176]{Janice~C.~Lee}
\STScI
\author{Rachel E. Pauline}
\UMICHD
\UIOWA
\author{Zion L. Thomas}
\CWRU
\author{William J. Berschback}
\OHHS
\author[0000-0001-7448-1749]{Kiana~F.~Henny}
\UWyoming
\author[0000-0002-5782-9093]{Daniel~A.~Dale}
\UWyoming
\author[0000-0002-0560-3172]{Ralf S.\ Klessen}
\ITA
\IWR
\author[0000-0002-3933-7677]{Eva~Schinnerer}
\MPIA
\author[0000-0002-3247-5321]{Kathryn~Grasha}
\altaffiliation{ARC DECRA Fellow}
\affiliation{Research School of Astronomy and Astrophysics, Australian National University, Canberra, ACT 2611, Australia}   
\AthreeD 
\author[0000-0003-0946-6176]{M\'ed\'eric~Boquien}
\UTA
\author[0000-0003-3917-6460]{Kirsten~L.~Larson}
\STScIESA
\author[0000-0003-1943-723X]{Sinan~Deger}
\Caltech
\author[0000-0003-0410-4504]{Ashley~T.~Barnes}
\Bonn
\author[0000-0002-2545-1700]{Adam~K.~Leroy}
\OSU
\author[0000-0002-5204-2259]{Erik~Rosolowsky}
\UAlberta

\author[0000-0002-0012-2142]{Thomas~G.~Williams}
\author[0000-0001-7130-2880]{Leonardo~\'Ubeda}
\STScI
\affiliation{Sub-department of Astrophysics, Department of Physics, University of Oxford, Keble Road, Oxford OX1 3RH, UK}

\begin{abstract}
We present new catalogs of likely globular clusters (GCs) in 17 nearby spiral galaxies studied as part of the PHANGS-HST Treasury Survey. The galaxies were imaged in five broad-band filters from the near-ultraviolet through the $I$ band.  PHANGS-HST has produced catalogs of stellar clusters of all ages by selecting extended sources (from multiple concentration index measurements) followed by morphological classification (centrally concentrated and symmetric or asymmetric, multiple peaks, contaminant) by visually examining the V-band image and separately by a machine-learning algorithm which classified larger samples to reach fainter limits. From both cluster catalogs, we select an initial list of candidate GCs to have $B-V \geq 0.5$ and $V-I \geq 0.73$~mag, then remove likely contaminants (including reddened young clusters, background galaxies misclassified by the neural network, and chance superpositions/blends of stars) after a careful visual inspection.  
We find that $\approx86$\% of the color-selected candidates classified as spherically symmetric, and 68\% of those classified as centrally concentrated but asymmetric are likely to be GCs.  The luminosity functions of the GC candidates in 2 of our 17 galaxies, NGC\,628 and NGC\,3627, are atypical, and continue to rise at least 1~mag fainter than the expected turnover near $M_V \sim -7.4$. These faint candidate GCs have more extended spatial distributions than their bright counterparts, and may reside in the disk rather than the bulge/halo, similar to faint GCs previously discovered in M101. These faint clusters may be somewhat younger since the age-metallicity degeneracy makes it difficult to determine precise cluster ages from integrated colors once they reach $\approx1$~Gyr.

\end{abstract}

\section{Introduction} \label{sec:intro}
\par

Globular clusters (GCs) are ancient, gravitationally bound stellar systems that are amongst the oldest luminous objects in the universe. They provide important insight to a broad range of fields in astrophysics, including stellar and chemical evolution \citep{VanLoon10}, clues to the star formation and assembly histories of galaxies \citep{Forbes18}, constraints on the epoch of reionization \citep{Moran14,Salvador-Sol17,Xiangcheng21}, the role of dark matter in structure formation \citep{Kruijssen19, Reina-Campos23}, and the distribution of dark matter in present-day galaxies \citep{Hughes21}. 

The Hubble Space Telescope (HST) has revolutionized our understanding of globular cluster systems in elliptical galaxies because early type galaxies generally have simple, relatively dust-free morphologies and their cluster populations are predominantly old (though limited star formation can occur; \cite{Sedgwick21}).
The GC systems in nearly all ellipticals have a log-normal distribution with a peak near $M_V \approx -7.4$ \citep{Villegas10, Jordan07}. 
The dispersion or width of the globular cluster luminosity function (GCLF) in ellipticals depends on the total magnitude of the host galaxy, with fainter galaxies having narrower smaller widths \citep{Jordan06, Jordan07}.

Our understanding of globular cluster systems in spiral galaxies remains far behind that in elliptical galaxies, since ongoing, higher intensity star formation creates complex structures that limit our ability to separate ancient clusters from the many younger clusters that also form in these galaxies, even with HST imaging.  The best-studied globular cluster system in a spiral galaxy is that in the Milky Way. The GCLF in the Milky can be described by a log-normal distribution with a peak at $M_V = -7.29 \pm 0.13$ and a dispersion parameter of $\sigma = 1.1 \pm 0.1$ mag \citep{Secker92}. This shape is similar to that found in ellliptical galaxies, but
different from the luminosity and mass functions of open clusters in the Milky Way and massive young clusters in other spiral galaxies, which are well-described by a power-law (e.g., M83: \citealp{Bastian12}; \citealp{Chandar14}, M51: \citealp{Chandar16}, Antennae: \citealp{Whitmore99}). The difference between the luminosity/mass functions of young and ancient clusters is believed to arise from the preferential destruction of lower mass clusters over time, likely due to evaporation driven by two-body interactions within clusters, and also bulge/disk shocking as the cluster orbits the host galaxy (e.g., \citealp{Fall01}). 

Characterizing the GC systems in spiral galaxies beyond the Local Group requires multi-band imaging with HST.  The shape of the luminosity functions in the few spiral galaxies with well-studied GC systems 
are similar to those in early-type galaxies, with a peak $M_V \approx -7.4$~mag \citep{Lomeli-Nunez22}.  
M101 is the single known exception, where the luminosity function of GC candidates continues to rise in a power-law fashion well below the expected peak \citep{Barmby06, Simanton15}.  Currently, little is known about the faint end of the GC luminosity function in spiral galaxies in general, since identifying even bright GCs and separating them from contaminants such as reddened young clusters and background galaxies can be challenging in these actively star-forming systems.

The PHANGS-HST survey \citep{Lee22} recently imaged 38 spiral galaxies in five broad-band filters and has complementary 
CO 2-1 maps from ALMA\citep{Emsellem22,Leroy21a}, and 2-21$\mu$m imaging for half the sample from JWST \citep{Lee23}  The collaboration has invested enormous effort to produce state-of-the art catalogs of $\sim$40,000 compact clusters of all ages \citet{Lee22,Thilker22,Deger22,Wei20,Whitmore21,Hannon23} (Maschmann et al, in prep.).  These catalogs provide an ideal starting point to identify and study ancient globular cluster populations in spiral galaxies.  In this work, we begin by investigating the globular cluster populations of 17 PHANGS-HST galaxies.

The rest of this paper is organized as follows. In $\S2$, we summarize the data, photometry, and the pipeline that was used to produce the initial compact cluster catalogs. In $\S3$ we select an initial list of globular cluster candidates based on their colors, and describe the subsequent visual inspection of each candidate to assess if it is likely to be a globular cluster or a contaminant. 
In $\S4$, we present our new catalogs of ancient globular clusters in 17 PHANGS-HST spiral galaxies and assess the contamination rates from the initial cluster selection.
In $\S5$ we present the luminosity functions of our GCs, identify excess, faint populations 
in two of the 17 galaxies and investigate what they might be, and in $\S6$ we state our conclusions and future studies using our new catalogs.

\begin{table*}
\centering
\caption{Properties of Sample Galaxies }
    \begin{tabular}{l c c c c c l}
        \hline
            Galaxy Name     & Distance  & Method & Reference & log$M_*$    & $SFR_{\rm total}$ & Morphology \\
                     &  (Mpc) & &    & (M$_{\odot}) $ & (M$_{\odot}$\,yr$^{-1}$) &  \\
        \hline    
        ~NGC\,628   &	9.84$\pm$0.63 & TRGB & 1 & 10.2 & 1.75 & SAc \\
        ~NGC\,1433  &  15.17$\pm$1.52 & PNLF & 5 & 10.4 & 1.13 & SBab \\
        ~NGC\,1559  &	19.44$\pm$0.44 & Mira & 6 & 10.2 & 3.76 & SBcd \\
        ~NGC\,1566  &  17.69$\pm$2.00 & Group & 4 &  10.7 & 4.54 & SABb \\
        ~NGC\,1672  &  19.40$\pm$2.91 & NAM & 2+3 & 10.6 & 7.60  & Sb\\
        ~NGC\,1792  &  16.20$\pm$2.43 & NAM & 2+3 & 10.5 & 3.70 & SAbc \\
        ~NGC\,2775  &  23.15$\pm$3.47 & NAM & 2+3 & 11.1 & 0.87  & SAc\\
        ~NGC\,3351  &  9.96$\pm$0.33  & TRGB & 1 & 10.3 & 1.32 & Sb \\
        ~NGC\,3627  &  11.32$\pm$0.48 & TRGB & 1 & 10.7 & 3.84 & SABb\\
        ~NGC\,4303 &  16.99$\pm$3.04 & Group & 4 & 10.6 & 5.33 & Sc\\
        ~NGC\,4321 &  15.21$\pm$0.49 & Cepheid & 7 & 10.7 & 3.56 & SABbc\\
        ~NGC\,4535 &  15.77$\pm$0.37 & Cepheid & 7 & 10.5 & 2.16 & SABc \\
        ~NGC\,4548 &  16.22$\pm$0.38 & Cepheid & 7 & 10.7 & 0.52 & SBb\\
        ~NGC\,4571 &  14.90$\pm$1.20 & Cepheid & 8 & 10.0 & 0.29 & SAd \\
        ~NGC\,4654 &  21.98$\pm$1.16  & Group & 7 & 10.5 & 3.79  & SABcd\\
        ~NGC\,4826 &  4.41$\pm$0.19  & TRGB & 9 & 10.2 & 0.20 & SABa\\
        ~NGC\,5248 &  14.87$\pm$1.34 & Group & 4 & 10.3 & 2.29 & SABbc\\
        \hline    
    \label{TAB:gal_properties}
    
    \end{tabular}
\begin{minipage}{12cm}

\small  \textbf{Note:} Distance, stellar mass ($M_*$), and star formation rate (SFR) are taken from the compilation in \citet{Lee22}.  Distance measurements were originally compiled in  \citet{Anand21}. Methods used to derive distances are shown in column 3, and include Tip of the Red Giant Branch (TRGB), Planetary Nebula Luminosity Function (PNLF), Cepheid Variables, Mira Variables, Numerical Action Method (NAM), and Galaxy Group velocity. Citations: 1) \citet{Jacobs09}, 2) \citet{Shaya17}, 3) \citet{Kourkchi20}, 4) \citet{Kourkchi&Tully17}, 5) \citet{Scheuermann22}, 6) \citet{Huang20}, 7) \citet{Freedman01}, 8) \citet{Pierce94}, 9) \citet{Anand21}.

\end{minipage}

\end{table*}
\newpage
\section{Observations, Detection, \& Photometry} \label{sec:obs}
\par
\subsection{Galaxy Sample}

\begin{figure}
    \centering
    \includegraphics[width = \textwidth]{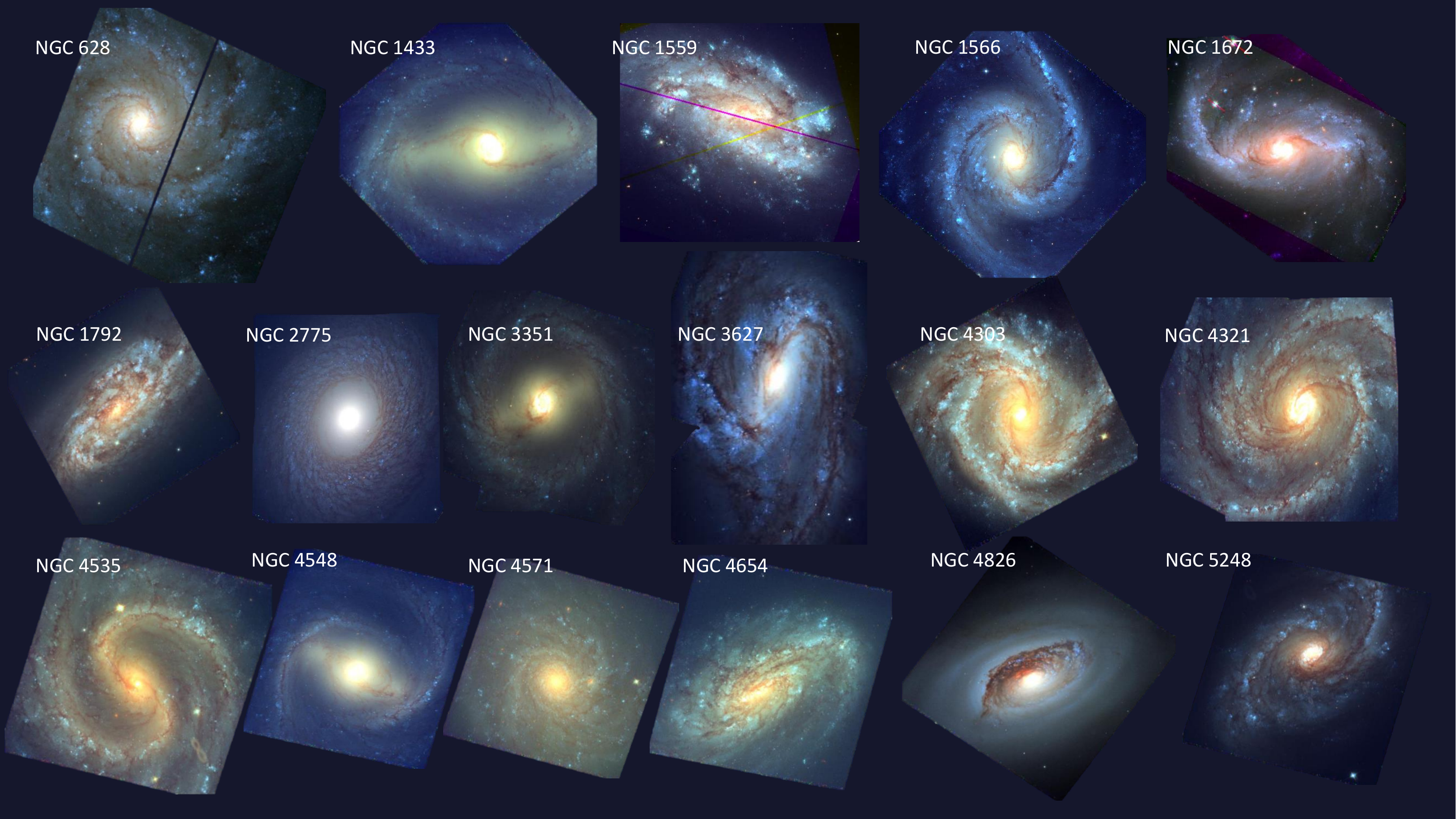}
    \caption{Color BVI images of the 17 PHANGS-HST galaxies studied in this work. The sample includes nearby face-on, massive spiral galaxies that cover a broad range in morphology, dust content, and spiral arm-type.
    }
    \label{fig:PHANGS_gal}
\end{figure}

Our sample includes 17 spiral galaxies from the PHANGS-HST sample \citep{Lee22}, which cover a large range in morphological classification (Sa through Sd, barred and unbarred), spiral arm type (flocculent and grand-design), total star formation rate (0.2 to 7.60 M$_\odot$ yr$^{-1}$), and mass ($\approx 10^{10}$ to $10^{11}$ M$_\odot$), representative of the present-day spiral population. 
All target galaxies are $\lea 20$~Mpc and fairly face-on, as indicated in the three-color HST images in Figure \ref{fig:PHANGS_gal}. 
Our sample includes nearly half of the entire PHANGS-HST sample of 38 galaxies.

We summarize the basic properties, including distance, total stellar mass ($M_*$), star formation rate (SFR), and morphological type in Table \ref{TAB:gal_properties}. Distance determinations are from the compilation in \citet{Anand21}, and are based on a variety of methods.
\citet{Anand21} determined new distances to four of the galaxies based on the tip of the red giant branch (TRGB). For the rest, they carefully assessed distance estimates in the literature, which can vary in quality. They gave the highest preference to measurements based on the TRGB method and Cepheid variable.  The next preference was for standard candle methods, such as the planetary nebula luminosity function (PNLF), surface brightness fluctations.  When no distance estimates based on stellar populations were available, modeling methods were used to infer distance, either from numberical modeling of their orbits within a group or from the Numerical Action Method model of \citet{Shaya17}, which is a non-linear model that attempts to reconstruct the 3D orbits of galaxies.

Star formation rates are based on total far ultraviolet luminosities from GALEX and infrared luminosities from WISE for each galaxy. The star formation rates were derived using the prescriptions laid out in \cite{Leroy19}. 
Estimates of stellar mass are derived from Spitzer IRAC 3.6$\mu m$ or WISE1 3.4$\mu m$ fluxes (when Spitzer is unavailable), using the prescription in \cite{Leroy21a}. 

\subsection{Observations, source detection, and photometry}


We start from an internal release version of the candidate cluster catalogs produced by the PHANGS-HST project (PI: J.C. Lee, GO-15654). 
For the full public cluster catalog release, see Maschmann et al. in prep (https://dx.doi.org/10.17909/jray-9798). 
PHANGS-HST is a Cycle 26 Treasury program, which obtained multi-band imaging of 38 nearby spiral galaxies from 2019-2021, in the F275W/($NUV$), F336W/($U$), F438W/($B$), F555W/($V$), and F814W/($I$) filters, taken primarily with the WFC3 camera, which has an $\approx 160 \times 160\arcsec$ field of view. Archival observations using the ACS/WFC camera were used when available. Here, we present results for 17 of these galaxies. 

Data reduction, source detection, and photometry are described in detail in \citet{Lee22}, \citet{ Thilker22}, and \citet{Deger22}. 
The processed images in each filter are in electrons~ $s^{-1}$ and aligned to a common grid with a scale of $0.04\arcsec~\mbox{pix}^{-1}$ , with North up and East to the left. The absolute astrometry of each image is determined from Gaia DR2 sources \citep{Gaia18,Lindegren18}. 
The pixel scale of $0.04\arcsec~\mbox{pix}^{-1}$ corresponds to a spatial sampling of 0.86~pc~pix$^{-1}$ in the nearest galaxy (4.41~Mpc) to 4.5~pc~pix$^{-1}$ in the furthest galaxy(23.15~Mpc). HST/WFC3 resolution varies with wavelength across the NUV-optical range, and with detector position, but is best ($\sim0.07-0.08\arcsec$ FWHM post-pixelation) at V-band, degrading by about 10$\%$ at NUV and I.

Sources were detected using a combination of DOLPHOT \citep{dolphot} and DAOstarfinder \citep{Thilker22}. To ensure all clusters (even the most extended) were recovered as sources, a secondary detection pass using photutils/DAOStarFinder with a slightly broaden kernel was conducted and the source lists were combined. These sources include individual stars and stellar clusters, compact associations, background galaxies and foreground stars. The ``all source" catalogs contain up to a few hundred thousand objects per galaxy. Aperture photometry was performed for each source using a 4-pixel radius aperture, which contains $\approx 50\%$ of the flux of a typical cluster at a distance of 12~Mpc, the median PHANGS-HST galaxy distance. The background level was determined in a 7--8 pixel annulus.
An empirical aperture correction is applied as described in detail by \citet{Deger22} and summarized by \citet{Thilker22}. Our photometry is in the VEGAMAG system. In the rest of this paper, we refer to the F275W filter as $NUV$, F336W as $U$ band, F435W or F438W as $B$ band, F555W as $V$, and F814 as $I$, for simplicity.

\subsection{Initial Selection PHANGS-HST of Cluster Candidates}

Initial cluster candidates were selected from the ``all source" catalogs largely based on their structural parameters and a magnitude cut, specifically a combination of concentration indices (CI), where CI is the difference in magnitude measured in 2 different apertures. Extended sources have larger values of CI than point sources. A total of 8 apertures were used to calculate two `multiple' concentration indices (MCI). The inner and outer MCI values are renormalized averages of the CI measured across different apertures: MCI$_{\rm inner}$ uses 4 apertures between 1 and 2.5 pixels in radius, and MCI$_{\rm outer}$ uses 4 apertures between 2.5 and 5 pixel radii. These CI were measured on synthetic clusters, which are generated from IMFIT \citep{Erwin15}, with an assumed Moffat profile with parameters chosen to represent the full range of cluster sizes and luminosities. 

A model cluster region in MCI
space was defined by the synthetic clusters (see \citet{Thilker22} for details), and applied to all detected sources to select an initial morphologically guided yet maximally inclusive catalog of potential clusters. Because the synthetic clusters were computed specifically for the distance of each galaxy, the model cluster region varied with distance. 

\subsection{Cluster Classification by Visual Inspection}


Each candidate cluster was carefully scrutinized in the V-band by co-author B. Whitmore, and classified into one of the following morphological types:

\begin{itemize}
    \item Class 1: Single-peaked, symmetric source with a radial profile more extended than a point source
    \item Class 2: Single-peaked, asymmetric source with a radial profile more extended than a point source
    \item Class 3: Compact stellar association, defined as having multiple peaks within a fixed radius 
    \item Class $\geq$4: Various contaminants (e.g. blends, background galaxies, pairs, triplets, foreground stars, etc.)
\end{itemize}

One key strength of this approach is the ability to identify and eliminate most background galaxies and chance superpositions/blends of two or three stars.  The main limitation is that a human can classify only a limited number of sources. Only the brightest $\approx 1000$ cluster candidates in each galaxy enter the ``human"-classified cluster catalogs, resulting in a faint magnitude limit which varies from galaxy to galaxy. The final sample of Class 1$+$2 human-classified sources includes 7867 total cluster candidates of all ages across our 17 galaxies (see also Table 1, Maschmann et al, in prep.) 

\subsection{Machine-Learning Cluster Classification}

The PHANGS-HST collaboration has also produced catalogs of cluster candidates that have been classified by machine learning (ML) algorithms \citep{Wei20, Whitmore21, Hannon23}. This approach has the advantage that more sources can be classified which reach fainter magnitudes. In this paper, we use the PHANGS machine learning candidate catalogs that were generated using the VGG neural network.

The VGG19-BN neural network classifies sources using the same morphological class system as the human-based catalogs. A key advantage of the ML algorithm is the ability to classify sources based on all available (NUV, U, B, V, and I) filters rather than just the V-band image.
Note that a single MCI polygon was applied to all galaxies to select the initial list of cluster candidates to be classified by the machine-learning algorithms, whereas this polygon was tuned to each galaxy (and therefore more restrictive) for the initial candidate list for human classification.

The neural network was trained on a set of objects, previously classified by co-author B. Whitmore, in 24 galaxies \citep{Hannon23} observed as part of the PHANGS-HST program, using 5-band imaging of the identified clusters. 
Objects classified using the VGG-BN machine-learning architecture matched the human classifications in PHANGS catalogs with $\approx 80\%$ accuracy in Class 1 and $\approx 50\%$ accuracy in Class 2 when tested over 5 PHANGS-HST galaxies\citep{Whitmore21}.

Overall, the machine learning selection adds 11,326 unique Class 1+2 candidate clusters to the human catalogs. The vast majority of these sources are fainter than the magnitude limits of the human-classified catalogs. 
A comparison of the human and ML-classified GC candidates above these  brightness limits show that the source lists contain the same objects at the $\approx80-90$\% level in nearly all galaxies,
with the ML-classified catalogs including a handful sources. NGC~1566 and NGC~1792 are exceptions, with the final ML catalogs including 105 and 104 more GC candidates, respectively, than the final human-classified catalogs discussed in Section~3.
The additional clusters in the ML catalogs are nearly all faint and close the magnitude limit of the human catalog.

After combining our human and ML classified catalogs at all magnitudes, we have a total of 19,193 candidate clusters of all ages across all 17 PHANGS-HST spiral galaxies.  The full human and machine-learning classified catalogs will be released in Maschmann et al., in prep.


\section{Production of Globular Cluster Catalogs}

One of the main goals of this work is to produce catalogs of ancient GCs with as few contaminants as possible, that are complete down to a known magnitude limit. While the PHANGS-HST pipeline initially estimates the ages of clusters using standard spectral energy distribution (SED) fitting techniques \citep{Adamo17, Turner21}, it has been shown that more than 70\% of ancient globular clusters can be incorrectly age-dated due to the age-reddening-metallicity degeneracy  for some galaxies \citep{Whitmore20, Whitmore23, Turner21, Hannon22}.  While work is in progress to improve the SED fitting procedure to address this issue, e.g. \citep{Thilker23}, here we select candidate GCs independently of the SED-fit ages.  Our catalogs can be used as a check on the results of improved SED fitting efforts.
We return to this point in section 3.5.  Our strategy here is to apply a color-based initial selection to the human- and machine-learning catalogs described above to separate ancient GC candidates from the many younger clusters in our spiral galaxy sample, and then to perform a visual inspection of all sources which pass the color selection to identify and reject remaining contaminants. 
It is important to note that broad-band colors of candidate GCs are subject to the age-metallicity degeneracy, which makes it quite challenging to distinguish clusters with ages $\approx 1$~Gyr from those with ages $\approx10$~Gyr based on photometry alone.  For this reason, we do not attempt to estimate the ages of our candidate GCs beyond identifying those which are $\approx1$~Gyr and older.  For our purposes, candidate globular clusters have all of the morphological and integrated photometric properties expected of globular clusters in the Milky Way, meaning they are centrally concentrated, round, and extended (broader than the PSF), have uniform red colors, and are not associated with ISM (dust, warm ionized gas, or molecular gas).

\subsection{Initial Color Cuts}

\begin{figure}
    \centering
    \includegraphics[width = \textwidth]{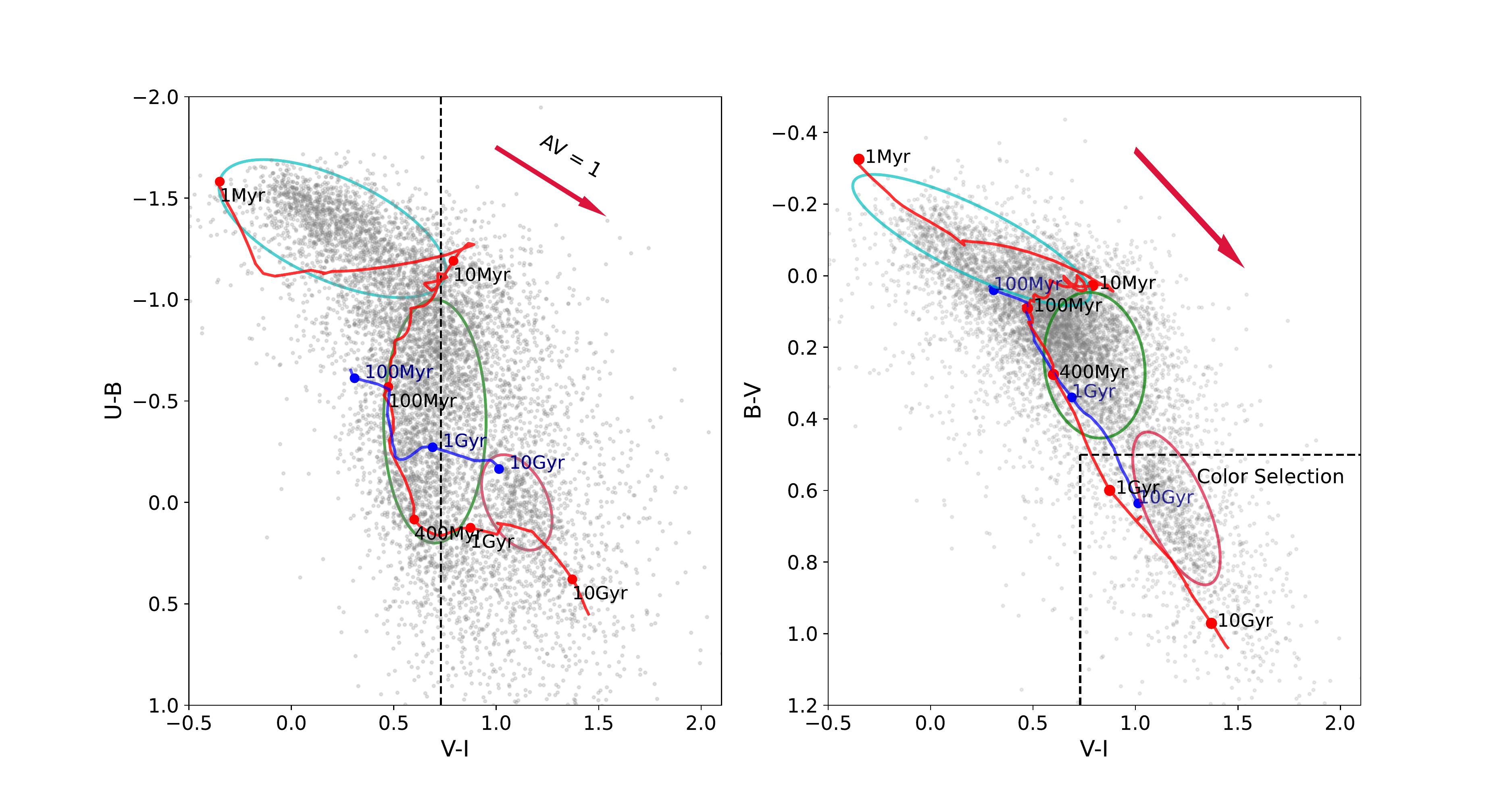}
    \caption{Color-color diagrams are shown for all human-selected Class 1$+$2 cluster candidates. 
        $U-B$ vs $V-I$  is shown on the left and $B-V$ vs $V-I$ on the right. 
    The predicted color evolution from the \citet{Bruzual03} stellar population models are shown for solar metallicity (red) and 1/50$\times$solar metallicity (blue). The light blue oval at the upper left shows the 'young cluster locus', the green oval shows the 'middle-aged plume', and the red oval shows the 'old globular-cluster clump'. 
    Dashed lines indicate the color cuts we use to select our initial list of GC candidates, B-V $\geq 0.5$ and V-I $\geq$ 0.73. Only those candidates with significant U-band measurements are plotted in the left panel. 
    The arrows represent the direction and amount of reddening for $A_V = 1$.}
    \label{fig:human12_cc}
\end{figure}

In Figure \ref{fig:human12_cc} we show two color-color diagrams of all Class 1+2 sources from the human-selected PHANGS-HST catalogs. With such large samples, color-color diagram provide a key reference for stellar, cluster, and galaxy evolution studies (Lee 2023 in prep). 
The left panel shows $U-B$ vs $V-I$ colors, while the right panel shows $B-V$ vs $V-I$. The diagrams include predictions from the solar metallicity (red) and the 1/50th solar (blue) evolutionary models of \citet{Bruzual03}, henceforth BC03. The models predict the intrinsic colors of extremely young (1~Myr) clusters to start at the top-left of each diagram and to become redder as they age and move toward the bottom right ($\approx10$~Gyr). The models do not include line or continuum nebular emission, and do not allow for binaries or rotation. The different metallicity models begin to diverge after a few 100~Myr, with the 1/50th solar model having bluer colors than the solar metallicity model at the same age due to the well-known age-metallicity degeneracy.

As discussed in Lee et al. 2023 in prep. and Maschmann et al. 2023 in prep, there are three distinct features in the color-color diagram: the young cluster locus, the middle-aged plume, and the old globular cluster clump. 
Starting in the upper left, in line with the youngest ages predicted by the BC03 evolutionary model tracks, we see the ``young cluster locus" ($< 10$~Myr, {\bf blue contour}). These very young clusters experience a range of reddening, from very little up to a few magnitudes in the $V$-band. They follow the slope of the reddening vector shown as the red arrow in each panel.
The ``middle-aged plume" highlights intermediate age clusters ($\simeq$ 30-400~Myr, {\bf green contour}), which hug the curved portion of the solar metallicity evolutionary model. These tend to have little reddening. Finally, we see a distinct over-density of clusters near the 10~Gyr mark of the 1/50th solar metallicity BC03 model, distinct from the intermediate age sources, with $V-I \approx 1.0 - 1.4$ and $U-B$ between $-$0.2 and 0.5~mag. Many of these are the ancient globular clusters we are interested in for this work. The bluest portion of this population falls quite close to 12~Gyr clusters predicted by the 1/50th solar metallicity model. An important point is this ``old globular cluster clump" ({\bf red contour}) separates from intermediate-age clusters in both color-color diagrams, although not as clearly in $B-V$ vs $V-I$ (Maschmann et al, in prep.)

To identify an initial list of candidate GCs, we select all Class 1$+$2 sources from the human and ML-selected catalogs that have $B-V \geq 0.5$~mag and $V-I \geq 0.73$~mag. This selection is indicated by the dashed lines in Figure \ref{fig:human12_cc}. We do not the N$U$V or $U$ bands in our color selection because GCs are fainter at bluer wavelengths and hence have larger photometric uncertainties in these bands. Our color cuts are consistent with the intrinsic colors of Galactic GCs\citep{Harris96, Harris10}.
Our color selection yields 1,262 (2,305) GC candidates from the human-classified (machine-learning classified) catalogs. These 3,567 sources form our initial list of GC candidates.

\subsection{Visual Inspection of GC Candidates}

For each GC candidate, we perform a visual inspection using several diagnostic plots and three-color images to assess:

\begin{itemize}
    \item \textbf{Morphology:} Is the source a single object, a close pair, or a compact association with multiple peaks? While restricting the selection to Class 1+2 (extended) sources eliminates many contaminants, previous works have shown that some remain \citep{Hannon22, Whitmore23}. 
    \item \textbf{Colors Relative to Model:} We use both the $U-B$ vs $V-I$ and $B-V$ vs $V-I$ color-color diagrams and predictions from the BC03 evolutionary models as a way to check that the colors of each cluster are reasonable. 
    \item \textbf{Presence or Absence of ISM:} We make use of archival HST-H$\alpha$ images for NGC~628, NGC~1433, NGC~1672, and NGC~3351 and ALMA CO maps for all galaxies to assess whether a candidate is associated with nearby ISM. The presence of ISM near a cluster indicates that it is likely to be young and reddened, rather than old with little reddening \citep[see also][]{Thilker23}. 
    \item \textbf{Local Environment:} Is the cluster isolated, or in a crowded star-forming region? Is it in the spiral arms, between them, or in the outskirts?  Candidates in crowded star-forming regions are more likely to be reddened, young clusters than ancient GCs. 
\end{itemize}

Each of these properties offers its own set of challenges and limitations. \textbf{For the cluster catalogs described in Section~2.4 which are based on visual classification, our independent classification process provides an important check which minimizes bias and unconscious shifting of selection criteria.}
Next, we show three representative examples from our initial GC candidate list.

\subsubsection{Example 1: Good Globular Cluster} 
\begin{figure}[h]
    \centering
    \includegraphics[scale = 0.38]{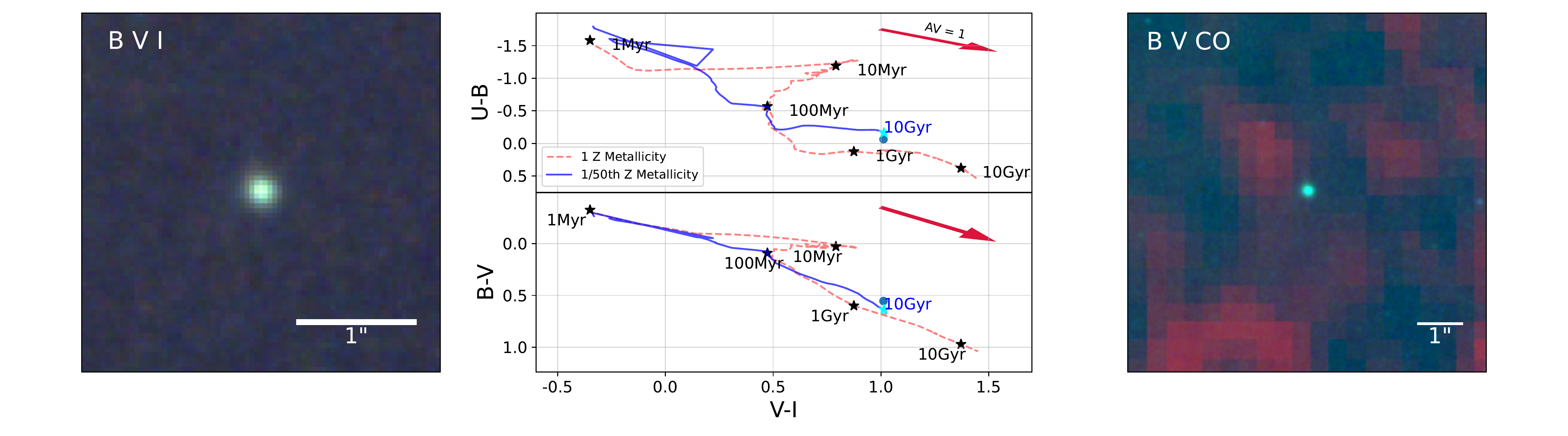}
    \caption{Diagnostic plots for an ancient globular cluster. The source appears symmetric, round, and has a uniform color in a $BVI$ image (left). The right panel, a $B-V-$~CO image, shows there is no CO emission associated with the cluster. In the color-color diagrams shown in the center panel, the cluster (blue circle) lies close to predictions from the $1/50 \times$solar metallicity SSP model.
    Black star symbols mark ages of 1~Myr, 10~Myr, 100~Myr, 1~Gyr, and 10~Gyr on the solar metallicity track.  A cyan star marks the predicted position for 10~Gyr in the 1/50th solar model.} 
    \label{fig:good glob}
\end{figure}

In Figure \ref{fig:good glob} we present an example of a likely globular cluster that passed our visual inspection. The left (right) panel shows postage stamp images in $B-V-I$ ($B-V-$~CO). The $B-V-I$ image shows the object is circular and centrally concentrated, that it has a `soft' appearance because it is broader than the PSF, and it has a uniform color. The $B-V-$~CO panel on the right shows there is no CO emission associated with the cluster, a strong indication that it is not reddened.  The colors of the cluster are quite similar to those predicted for a metal-poor, ancient stellar system.

In general, since GCs are old and have red broad-band colors, they can be quite faint in the bluer filters. Therefore, we expect them to have higher photometric uncertainties in the $NUV$ and $U$ bands than in $B$, $V$, and $I$. While we use the $U$ band as a check in cases where the candidate is sufficiently bright, we do not eliminate clusters based on their $U-B$ color alone. The key takeaways from our inspection of this object are its soft, spherical appearance
and that it is not associated with any ISM. These clues lead us to conclude with high confidence that this is a bona fide ancient globular cluster.

\begin{figure}[h]
    \raggedleft
    \includegraphics[scale = 0.38]{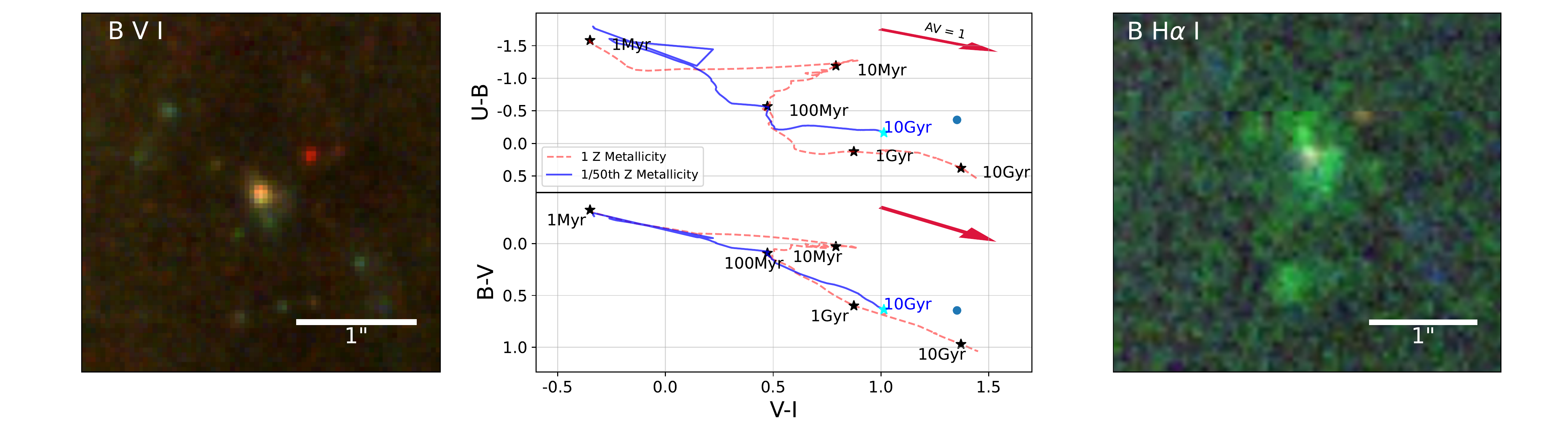}
    \caption{Diagnostic plots for a reddened young cluster. In the middle panels the source occupies a similar region in color-color space as a GC, but it is in a somewhat crowded region, appears lumpy, and has a non-uniform color in a $BVI$ image (left). In the right panel ($B-{\rm H}_\alpha-I$) there is a partial bubble of H$\alpha$ emission encircling the cluster. These are strong indications that the source is young and reddened rather than ancient. The color-color diagrams are in the same format as Figure \ref{fig:good glob}}
    \label{fig:Interloper Dust Lane}
\end{figure}

\subsubsection{Example 2: Reddened Young Cluster} In Figure \ref{fig:Interloper Dust Lane} we present diagnostic plots for a reddened young cluster, the type of source expected to be the largest source of contamination in our candidate GC catalogs. Whereas their broadband colors can be similar to GCs, they tend to be in crowded star-forming regions and have associated H$\alpha$ and/or CO emission. The left panel ($BVI$ three-color image) shows the source has a non-uniform color, and is not completely symmetric. In the right panel we see that it is surrounded by a bubble of H$\alpha$ emission, a clear indication it is a very young object. While this object's colors are similar to those of a GC in $B-V$ and $V-I$, its $U-B$ color is somewhat bluer than predicted for ancient clusters.
A reddening vector equivalent to $A_V=1$~mag in each color-color plot shows that the cluster could potentially intersect the evolutionary models at different ages, depending on the amount of reddening. The H$\alpha$ emission associated with this particular candidate leads us to conclude, with high confidence, that this is a reddened young cluster and not a GC. Even without H$\alpha$ emission, one can see the source is experiencing differential reddening (left panel), since the bottom edge is visibly less red than the majority of the source.

\begin{figure}
    \centering
    \includegraphics[scale = 0.38]{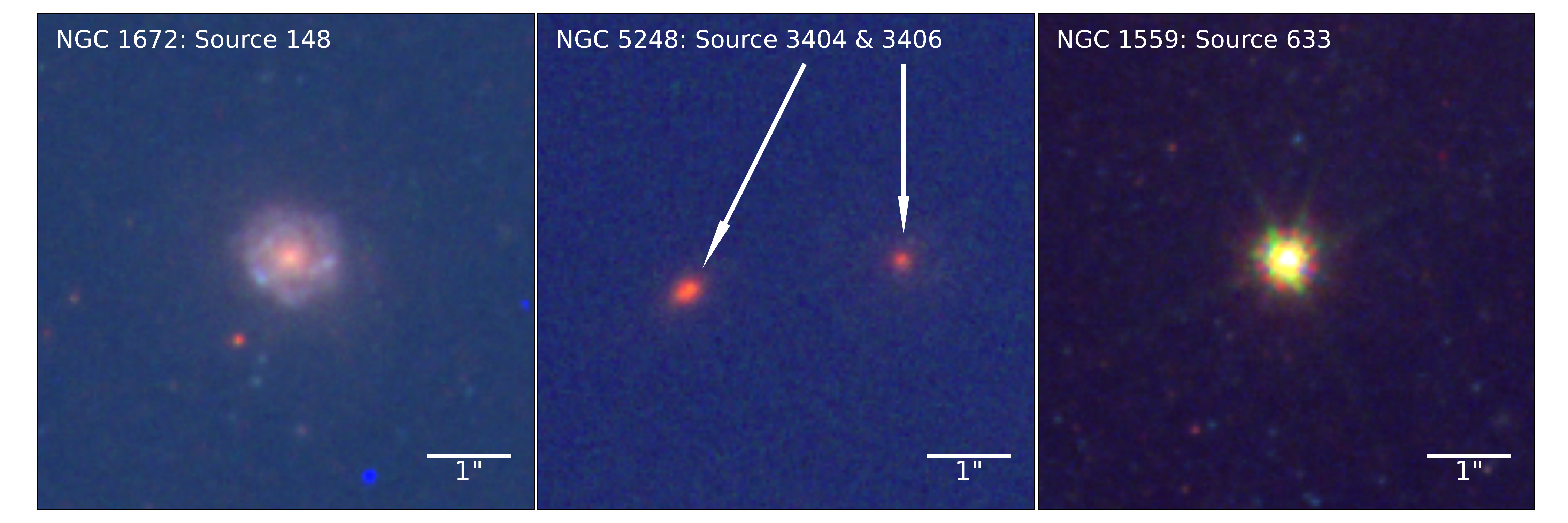}
    \caption{Color BVI images showing examples of contaminants from our initial color-selection of GC candidates. The left image shows a background galaxy that was included in the machine-learning catalogs. The center panel shows two background galaxies also from the ML-selected catalogs, while the right panel shows a saturated foreground star from the human-classified catalog. All contaminants are removed from the final GC catalogs.}
    \label{fig:RG_FGS}
\end{figure}

\subsubsection{Example 3: Background Galaxies and Foreground Stars} Background galaxies make up a small percentage of the contaminants in the initial selection of 3,567 GC candidates, and are almost exclusively found in the ML-based catalogs. While GCs and background galaxies both have extended profiles, galaxies often have faint additional features such as spiral arms, disks, or extended halos, which are fairly easy to identify for humans but can be challenging for the ML algorithm. The left and center panel of Figure \ref{fig:RG_FGS} shows 3 examples of background galaxies which made it into the ML-selected catalogs. The galaxy in the left panel shows clear spiral structure while the two in the center panel are identifiable by their ellipticity and extremely red colors, $V-I\geq 2.1$.  These and other similar objects are all removed from our GC final sample.

Foreground stars are rare in our initial candidate GC list since most were eliminated by the MCI cuts.
The right panel of Figure~\ref{fig:RG_FGS} shows an example of a foreground star with (weak) diffraction spikes which did make it through the human classification. This can happen occasionally since bright foreground stars are not necessarily saturated in every filter and can therefore slip through the initial candidate selection.  These objects are also removed from our final GC catalogs.

\section{Final Globular Cluster Catalogs}

\subsection{Human-Classified Catalogs}

\begin{table}[ht]
\caption{Statistics for Bright GC Candidates}
	\label{HUMAN_GC_TAB}
	\centering
	\begin{tabular}{l|cccccccc}
	    \hline\hline
		Galaxy   & Cand.\ & Conf.\  & \% Conf.\ & Cand.\ & Conf.\ & \% Conf.\ & Total & Mag.\ Lim.\ \\
		         &  H  GCs    & H GCs  &            & ML GCs & ML GCs &  & GCs   &  m$_V$\\
           (1) & (2) & (3) & (4) & (5) & (6) & (7) & (8) & (9) \\
		\hline \hline
        ~NGC\,628 &    82     &   67  &    81.7\%  &  88 & 70 & 83.8\% & \textbf{70} &23.0\\
		\hline
        ~NGC\,1433 &    53     &  51   &    96.2\%  &  58 &  53 & 96.3\% & \textbf{54} &24.1\\
        \hline
        ~NGC\,1559 &    89     &  63   &    70.8\%  &  105 &  71 & 74.2\% & \textbf{73}  &23.5\\
        \hline
        ~NGC\,1566 &    78    &  71   &    91.0\%  &  171 & 136 & 80.8\% & \textbf{139} &24.3\\
        \hline
        ~NGC\,1672 &    31     &  25   &    80.7\%  &  45  & 27 & 71.4\% & \textbf{27} &23.0\\
        \hline
        ~NGC\,1792 &    121    &  92  &    76.0\%  &  244 & 178 & 75.6\% & \textbf{181} &24.0\\
        \hline
        ~NGC\,2775 &    121    &  106  &    87.6\%  &  117  & 107  & 91.6\% & \textbf{115} &24.5\\
        \hline
        ~NGC\,3351 &    60     &  50   &    83.3\%  &  74  & 60  & 86.9\% & \textbf{62} &24.0\\
        \hline
        ~NGC\,3627 &    92     &  55   &    59.8\%  & 105  & 58  & 59.4\% & \textbf{61} & 22.5\\
        \hline
        ~NGC\,4303 &    23     &  15   &    65.2\%  & 25  & 15  & 71.4\% & \textbf{15} &22.5\\
        \hline
        ~NGC\,4321 &    76     &  58   &    76.3\%  & 99 & 67  & 71.9\% & \textbf{66} &23.5\\
        \hline
        ~NGC\,4535 &    57     &  47   &    82.5\%  & 68  & 56  & 83.3\% & \textbf{56} &24.0\\
        \hline
        ~NGC\,4548 &    83     &  73   &    88.0\%  & 93  & 81  & 87.5\% & \textbf{85} &24.5\\
        \hline 
        ~NGC\,4571 &    38     &  32   &    84.2\%  & 48  & 38  & 83.3\% & \textbf{41} &24.5\\
        \hline
        ~NGC\,4654 &    98     &  66   &    67.4\%  & 121 & 78  & 66.9\% & \textbf{80} &24.0\\
        \hline
        ~NGC\,4826 &    54     &  38   &    70.4\%  & 51  & 38  & 80.0\% & \textbf{42} &23.5\\
        \hline
        ~NGC\,5248 &   106     &  79   &    74.5\%   & 136 & 91  & 68.6\% & \textbf{94} &24.0\\
        \hline
        \hline
        Total/Mean$\%$ &  1262  &  988 &    78.3\%   & 1648 & 1225 & 74.3\% & \textbf{1261}

    \end{tabular}
	\tablecomments{Column 2 shows the number of color-selected GC candidates from the human catalogs, Columns 3 and 4 give the number and percentage of these that are visually verified or confirmed to be likely GCs. Columns 5, 6, and 7 give the analogous information for the machine learning-based catalogs. Column 9 shows the limiting magnitude for each galaxy.}

\end{table}

The results of our visual inspection of the initial color-selected GC candidate lists 
are compiled in Table \ref{HUMAN_GC_TAB}, and shown visually in Figure \ref{fig:bar_bright}. We include the initial number of color-selected GC candidates (Column 2), how many are visually verified as likely GCs (Column 3), and the percentage verified (Columns 4). The magnitude limit of the human-classified catalogs are listed in Column 9 for each galaxy. The initial color selection yielded between $20-120$ candidate GCs per galaxy, with most having 50 or more. The percentage of these that are confirmed by our rigorous visual inspection to be likely GCs ranges from $\sim 60-96\%$.   Out of 1,262 GC candidates, we visually confirmed that 988 appear to be likely GCs (78$\%$).

Figure \ref{fig:mosaic_GC_human} shows a representative sample of globular clusters from our final catalogs: the top two rows show examples of blue (presumably metal-poor) GCs and the bottom two show red (presumably metal-rich) GCs. We have included examples from grand design and from flocculent spirals. They all have a similar appearance: round, singly-peaked, and fairly isolated with very little nearby ISM.

Overall, we find that a simple color selection successfully identifies likely globular clusters at the $\approx 75-80$\% level from the full PHANGS-HST human-classified cluster catalogs, with no additional selection criteria.

\subsection{Machine Learning-Classified Catalogs}

\begin{figure}
    \centering
    \includegraphics[width = \textwidth]{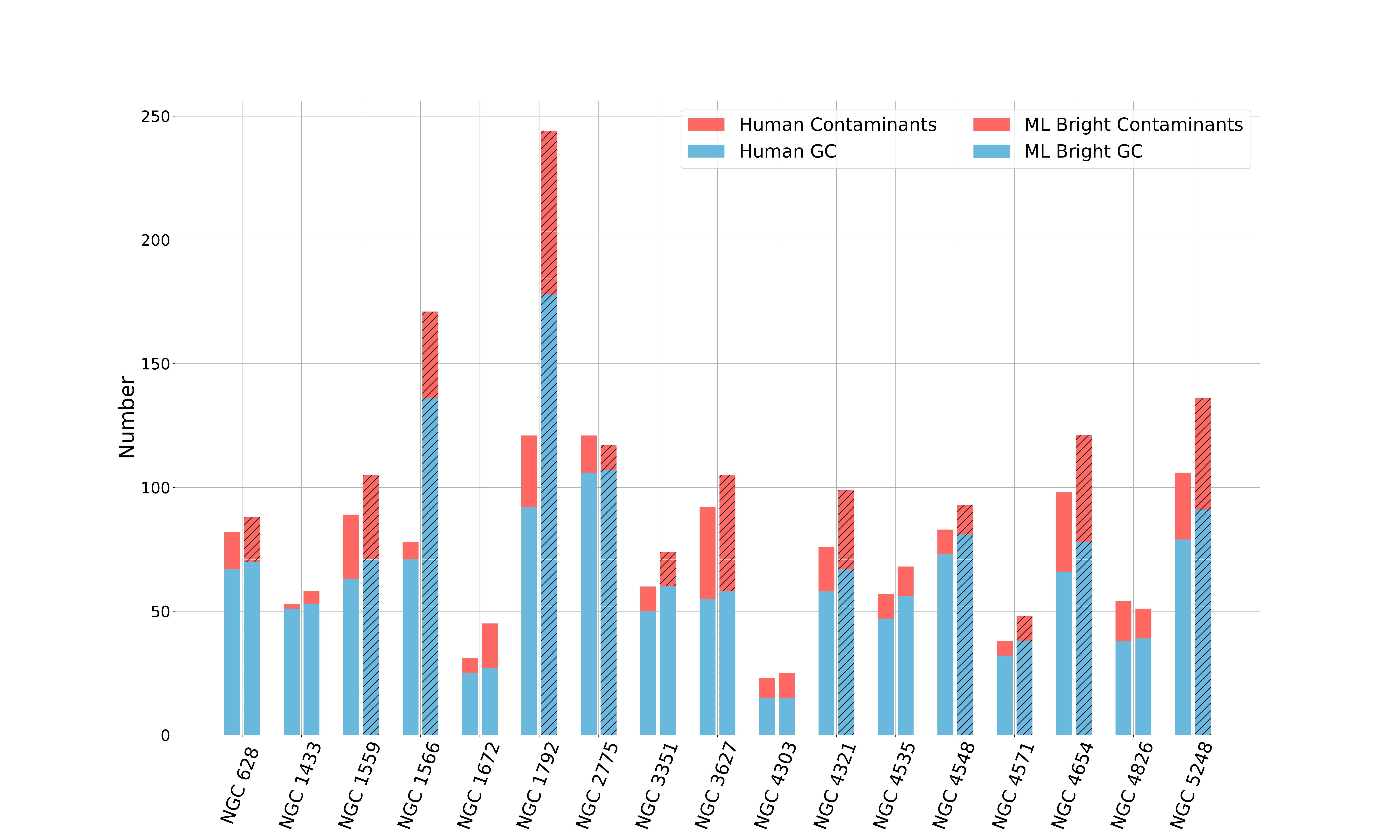}
    \caption{Total number of color-selected GC candidates from the human-classified catalogs (left bar) and the 'bright' (same magnitude range) ML-classified catalogs (right bar). Each bar is broken into likely GCs (blue) and contaminants (red).}
    \label{fig:bar_bright}
\end{figure}

\begin{figure}[h!]
    \centering
    \includegraphics[width = \textwidth]{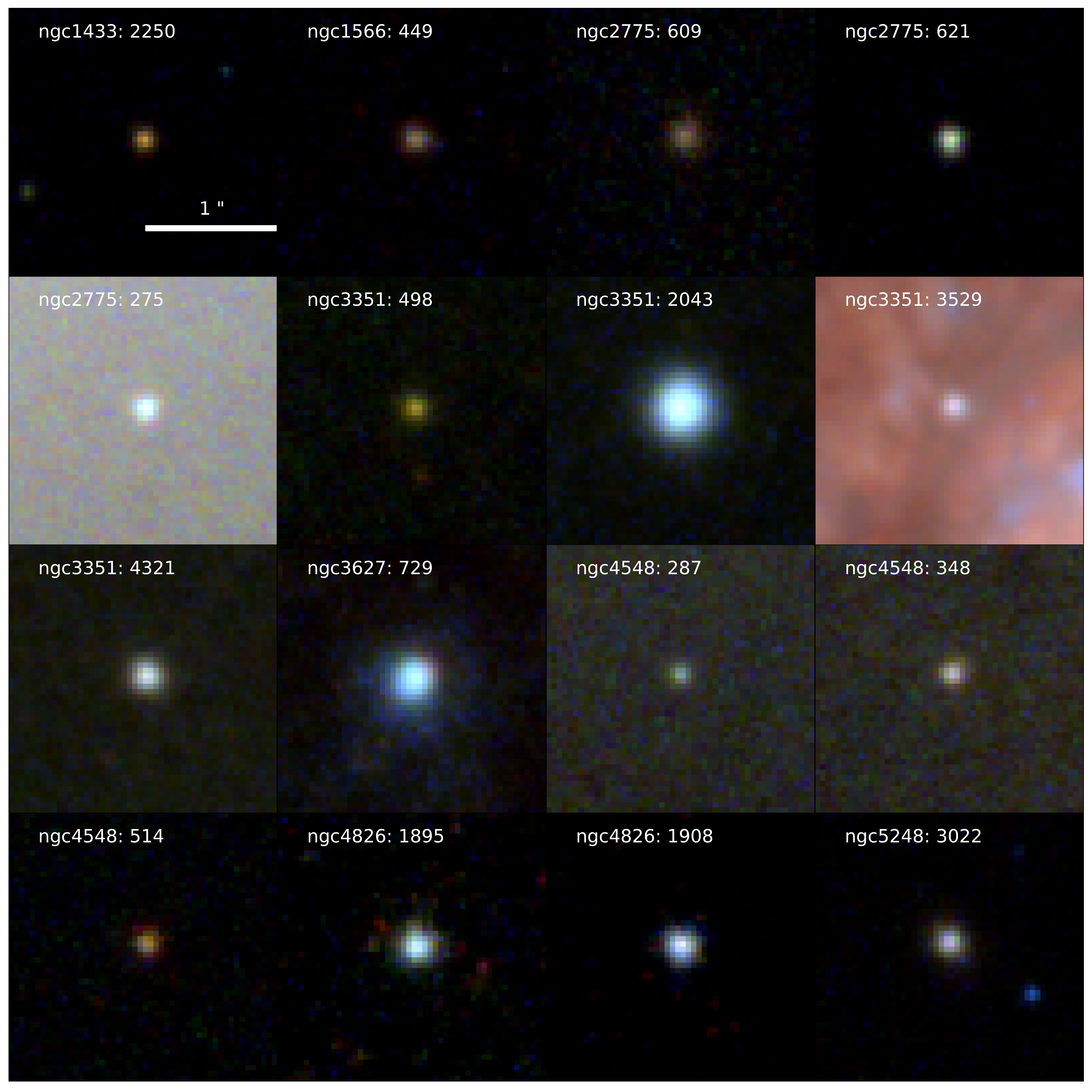}
    \caption{This figure shows BVI 2 $\times$ 2$\arcsec$ postage stamp images for 16 representative globular clusters from the human catalog. The clusters are all spherical and centrally concentrated, but show a range of colors.  Most are isolated with the exception of cluster 3529 in NGC~3351, which is found in the bulge.}        
    \label{fig:mosaic_GC_human}
\end{figure}

We assess two subsets of ML-classified GC candidates: brighter and fainter than the limiting magnitude listed in Column 9 of Table \ref{HUMAN_GC_TAB}. The number of GC candidates initially selected per galaxy was $\sim50-240$, with $\sim 60-96\%$ confirmed. The subset of `bright' cluster candidates consists of 1,648 sources. We confirm 1,225 of these from visual inspection, an overall success rate of 74.3$\%$. The statistics of color-selected `bright' sources and the subset of likely GC candidates as well as $\%$ confirmed in each galaxy are compiled in Columns~5-7 of Table~\ref{HUMAN_GC_TAB}, and shown in Figure \ref{fig:bar_bright}.

The number of confirmed GCs from our simple color selection from the human and ML samples are within $\approx 10\%$ of one another for the individual galaxies, with nearly all the same GC candidates found in the human and ML-based samples.
The only notable outliers are
NGC~1566 and NGC~1792, where the ML-based catalogs provide roughly double the number of likely GC candidates than the human-based ones.
The bright subsample of the ML-classified catalogs only yields a 4$\%$ difference in the average confirmed GCs when compared with the human-selected catalogs down to the same magnitude limit. 

\begin{table}[ht]
\caption{Statistics for Faint GC Candidates from ML Catalogs}
    \label{ML_GC_TAB_FAINTER}
	\centering
	\begin{tabular}{lccccc}
	    \hline\hline
		Galaxy & \# Candidate & \# GCs  &\% GCs & Mag.\ Range\\ 
		       &   GCs     &  & &m$_V$ (mag)\\
         (1) & (2) & (3) & (4) & (5)\\
		\hline \hline
        ~NGC\,628 &  225  &  166   & 70.5\%  & 23.0--24.9\\
		\hline
        ~NGC\,1433 &  46   &  40   & 85.4\%  & 24.1--25.0\\
        \hline
        ~NGC\,1559 &  120  &  100  & 80.0\%  & 23.5--25.1\\
        \hline
        ~NGC\,1566 &  42  &  37  & 82.0\%  & 24.3--24.9\\
        \hline
        ~NGC\,1672 &  339  &  266  & 75.4\%  & 23.0--25.8\\
        \hline
        ~NGC\,1792 &  15  &  15  & 95.8\%  & 24.0--24.3\\
        \hline
        ~NGC\,2775 &  $\cdots$  &  $\cdots$  & $\cdots$  &  $\cdots$ \\
        \hline
        ~NGC\,3351 &  50  &  43  & 81.5\%  & 24.0--25.3\\
        \hline
        ~NGC\,3627 &  564  &  460 & 78.3\%  & 22.5--24.7\\
        \hline
        ~NGC\,4303 &  131  &  79 & 60.0\%  & 22.5--24.4\\
        \hline
        ~NGC\,4321 &  201  &  172 & 77.7\%  & 23.5--24.8\\
        \hline
        ~NGC\,4535 &  17   &  16  & 91.7\%  & 24.0--24.3\\
        \hline
        ~NGC\,4548 &  1   &  1  & 100\%  & $\cdots$\\
        \hline 
        ~NGC\,4571 & $\cdots$   &  $\cdots$  & $\cdots$  & $\cdots$ \\
        \hline
        ~NGC\,4654 &  12  &  11  & 90.0\%  &  24.0--24.3\\
        \hline
        ~NGC\,4826 &  $\cdots $  &  $\cdots$  & $\cdots$  & $\cdots$ \\
        \hline
        ~NGC\,5248 &  47  &  38 & 78.3\%  &  24.0--24.7\\
        \hline
        \hline
        Total/Mean$\%$ &  1810  &  1444 &    79.8\%

	\end{tabular}
	\tablecomments{From the ML catalogs, we compile the total number of color-selected GCs (column~2), and the number and percentage confirmed through our visual inspection (columns~3 and 4) in the magnitude range given in column~5. }
	\end{table}

In Table \ref{ML_GC_TAB_FAINTER} we compile the results of our visual classification of `faint' GC candidates from the ML catalogs. These have the magnitude range given in Column 5. The `faint' subsample has a total of 1810 color-selected candidates, of which 1444 are confirmed as likely GCs. This gives an overall success rate of $79.8 \%$, very similar to that for the brighter human and ML-selected GC candidates. 

The final globular cluster catalogs created in this work can be found at https://dx.doi.org/10.17909/jray-9798.

\subsection{Overall Statistics}

\begin{table}[]
    \centering
    \caption{Total Number of GCs}
    \begin{tabular}{lllc} \hline \hline
		~~~~Galaxy & GC Candidates & Final GCs  & Mag.\ limit (m$_V$)\\
        ~~~~(1)   &  (2) (bright + faint)  &  (3)  &  (4)  \\
		\hline \hline		
        1~~~NGC\,628 & \textbf{315}(90 + 225)&  \textbf{236} (70 + 166) & 24.9\\
        \hline
        2~~~NGC\,1433 & \textbf{105}(59 + 46) & \textbf{94} (54 + 40)& 25.0\\
        \hline
        3~~~NGC\,1559 & \textbf{233}(113 + 120) &  \textbf{173} (73 + 100) & 25.1\\
        \hline
        4~~~NGC\,1566 & \textbf{217}(175 + 42) &   \textbf{176} (139 + 37) & 24.9\\
        \hline
        5~~~NGC\,1672 & \textbf{385}(46 + 339) &   \textbf{293} (27 + 266) & 25.8\\
        \hline
        6~~~NGC\,1792 & \textbf{267}(252 + 15) &   \textbf{196} (181 + 15) & 24.3\\
        \hline
        7~~~NGC\,2775 & \textbf{131}(131 + 0) &   \textbf{115} (115 + 0) & 24.5\\
        \hline
        8~~~NGC\,3351 & \textbf{130}(80 + 50) &   \textbf{105}  (62 + 43) & 25.3\\
        \hline
        9~~~NGC\,3627 & \textbf{681}(117 + 564) &   \textbf{521} (61 + 460) & 24.7\\
        \hline
        10~~NGC\,4303 & \textbf{160}(29 + 131) &   \textbf{94}  (15 + 79) & 24.4\\
        \hline
        11~~NGC\,4321 & \textbf{303}(102 + 201) &   \textbf{238} (66 + 172) & 24.8\\
        \hline
        12~~NGC\,4535 & \textbf{89}(72 + 17) &   \textbf{72}  (56 + 16) & 24.3\\
        \hline
        13~~NGC\,4548 & \textbf{99}(98 + 1) &   \textbf{86}  (85 + 1) & 24.5\\
        \hline 
        14~~NGC\,4571 &  \textbf{53}(53 + 0) &  \textbf{41}  (41 + 0) & 24.5\\
        \hline
        15~~NGC\,4654 &  \textbf{145}(133+ 12) &  \textbf{91} (80 + 11) & 24.3\\
        \hline
        16~~NGC\,4826 & \textbf{63}(63 + 0) &   \textbf{42}  (42 + 0) & 23.5\\
        \hline
        17~~NGC\,5248 &  \textbf{191}(144 + 47) &  \textbf{132} (94 + 38) & 24.7\\
        \hline \hline 
        Total & \textbf{3567} & \textbf{2705} \\
        \hline
    \end{tabular}
    \label{FINAL_GC_NUMBERS}
\end{table}

\begin{figure}
    \centering
    \includegraphics[width = \textwidth]{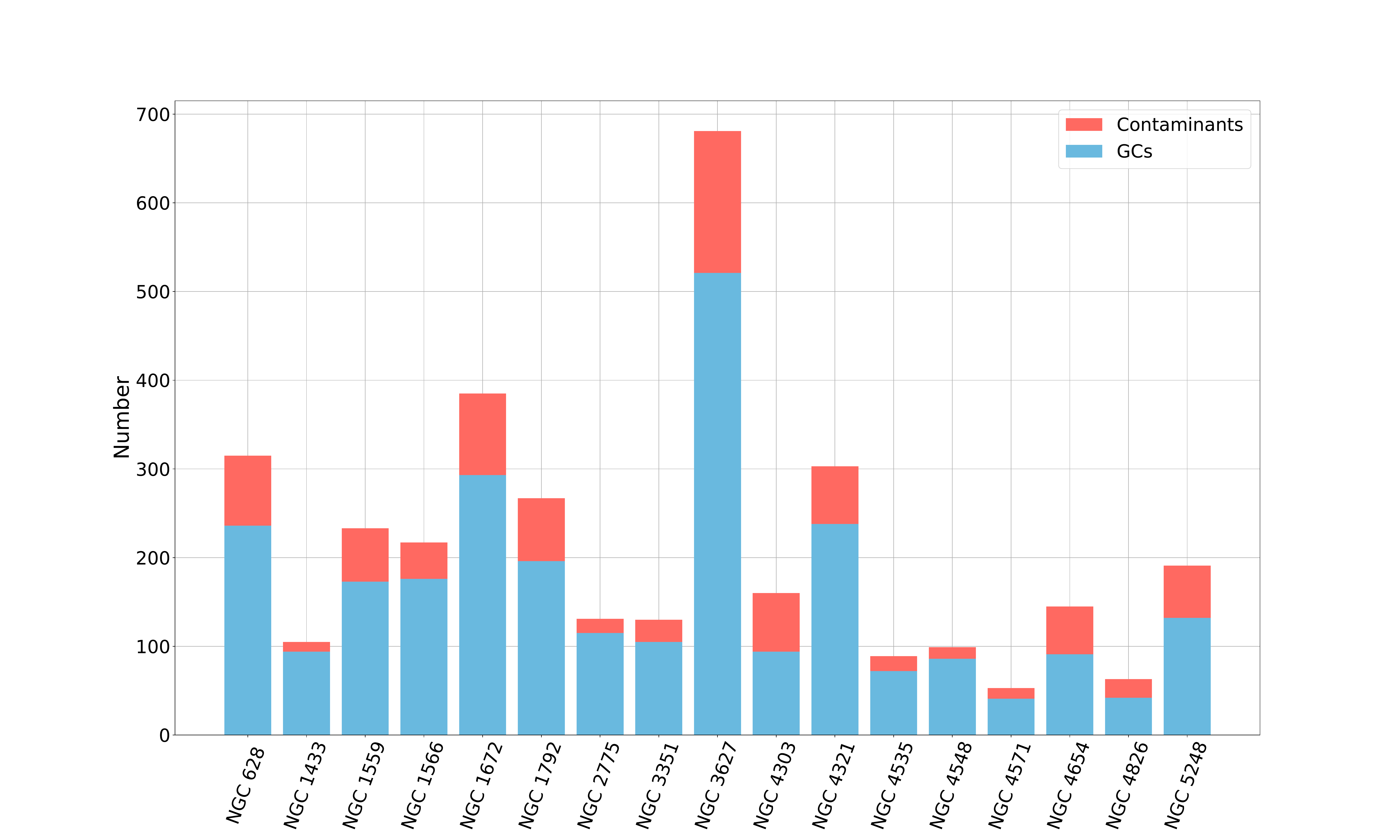}
    \caption{The total number of color-selected candidate GCs are shown for each galaxy in the sample. The contaminants are shown in red and likely GCs in blue. NGC~3627 is an outlier due to the large number of faint candidates selected by the ML algorithm.}
    \label{fig:All Source Bar Graph}
\end{figure}

 We started with 3,567 color-selected GC candidates, with 1,262 from the human-selected catalogs and 2305 from ML-selected ones. From our careful visual inspection, we confirm that 2,705 out of 3,567 are quite likely to be GCs, with 988/1,262 (78\%) from the human-based catalogs and 1,717/2,305 (75\%) from the ML-based catalogs. Table \ref{FINAL_GC_NUMBERS} lists the final numbers of initial candidates and confirmed GCs in each galaxy. Results for individual galaxies are displayed in Figure \ref{fig:All Source Bar Graph}.

The differences between objects classified as Class 1 vs Class 2 is notable here. GCs are dynamically relaxed systems, and therefore expected to be spherically symmetric, i.e. Class 1. We find that for Class 1, 1,332 out of 1,547 initial candidates are classified as likely GCs (a success rate of 86\%), whereas for Class 2, we find 1,373 out of 2,020 initial candidates are likely GCs (a 68\% success rate). 
The human-classified class 2 catalogs have a lower success rate (56\%) for likely GCs when compared with the ML catalogs (71\%). While the success rate of finding likely GCs based on a simple color-selection is quite high for Class~1 objects, 
Class~2 sources contribute a significant fraction of likely GCs as well, and our catalogs would be significantly less complete without them.  We conclude that for constructing GC catalogs, it is best to include both Class 1 and 2 sources.

\subsection{Checks \& Limitations}
\label{Limitation}
Of the 3,567 initial, color-selected GC candidates, we found above that 2,705 are retained in our final GC catalogs.  This means that  862 sources were eliminated through our visual inspection.  Of these, we believe that 589 (212 human/ 377 ML) are good clusters but are just younger and reddened (with ages $\approx 1 - 400$~Myr), while the remaining 273 (62 human/ 211 ML) are background galaxies or other artifacts.
In the rest of this section, we estimate the level of contamination that may still be present in our final GC catalogs from: reddened young clusters, background galaxies, foreground stars, and partially reddened intermediate-age sources. 

We find that {\bf reddened young clusters} are the dominant source of contamination in our color-selected GC candidate catalogs. Of the 862  contaminants that we have identified, 589 are younger clusters. Of these, 295 appear to be heavily reddened (median $E(B-V) \approx 0.7$, up to 1.5 mag) young clusters ($\tau < 10$~Myr), and 257 to be moderately reddened (median $E(B-V) \approx 0.5$ mag), intermediate age clusters ($10 < \tau < 400$~Myr). The remaining 37 appear to be lower mass clusters which have red colors due to 
stochastic sampling of the stellar IMF; we expect these clusters to be young with ages $\approx$ 10-20Myrs. Overall, reddened clusters account for $\approx60\%$ (589/862) of all sources removed during visual inspection, with the remaining 273 sources being non-cluster contaminants.  

Very young, reddened $\tau \leq 6$~Myr clusters can be identified by the presence of H$\alpha$ emission associated with the cluster see also Thilker et al.s in prep. Four galaxies in our sample (NGC\,628, NGC\,1433, NGC\,1672, and NGC\,3351) have archival HST-H$\alpha$ images available. We use these to estimate the number of reddened young clusters which may have been missed and remain in the final GC catalog for the other 13 galaxies.

To test this, we visually inspect the 216 human-selected GC candidates in the 4 galaxies which have HST $H_{\alpha}$, once making use of the H$\alpha$, and once without. Of the inspected candidates, we find 4 sources which appear to be reddened, young clusters that were incorrectly classified as a GC when the H$\alpha$ images were not used. This gives an expected misclassification rate of $\approx1.8$\% or $\approx60$ out of the 3351 remaining candidate GCs not included in this test.

Another type of contaminant we consider is {\bf partially reddened young or intermediate age clusters}. We identify 45 faint sources that appear to be only partially reddened by dust, where the cluster shows distinct red and blue colors. Because it is possible that this partial reddening has led clusters that are only $\simeq$ 100 Myr to be misclassified as ancient GCs, we compared the $U-B$ and $V-I$ colors of the bluest 5 pixels of the candidates to the BC03 models. We identified and eliminated just 1 source that appears to be a partially reddened, intermediate-age cluster rather than an ancient GC.

We identify non-cluster contaminants based on morphology and colors, particularly {\bf background galaxies}. We find 129 background galaxies, which comprise $47.3\%$ of all non-cluster contaminants. We expect any remaining contamination from background galaxies to be negligible in the final catalogs, since these objects are visually distinct from clusters. As shown in Figure \ref{fig:RG_FGS}, background galaxies are redder than GCs in most cases, and often have visible structure. The exception to this is E0 galaxies, which can have a similar morphology to GCs but with an extended faint red halo. 

{\bf Foreground stars} are usually removed from the cluster catalogs based on their concentration index, but a few made it through the initial selection criteria because they are (barely) saturated and therefore appear broader than the PSF.  Nearly all such cases have been eliminated from the human-classified catalogs, but a few made it through the machine learning-classified catalogs.

Besides misclassification, it is probable that we have missed some ancient GCs because they are in regions of high background (bulge or spiral arms), crowding, or dust obscuration. In particular, we expect our samples to become increasingly incomplete at fainter magnitudes and closer to the galactic center (we will discuss where we think incompleteness may impact the shape of the cluster luminosity function in Section~5.2).


After carefully examining thousands of candidate GCs selected from an initial color cut of $B-V \geq 0.5$ and $V-I \geq 0.73$~mag, we conclude that a number of remaining background galaxies can be eliminated by applying a cut at the red end. We found a number of background galaxies with $V-I >2.1$~mag, but essentially no GCs with such colors, at least in the galaxies studied in this work. We check the impact of applying a color cut with the following criteria: (i) $V$--$I$ $\leq 2.1$ and (ii) $B$--$V$ $\leq -0.217 \times \mbox{(V-I)} + 1.85$. These cuts removed 100 sources, of which 90 are background galaxies and 10 appear to be good GC candidates. If we had included these cuts in our initial selection, the confirmed percentages of GCs would only change by 1--2\% for most galaxies, but a number of contaminants could have been eliminated before visual inspection.

\subsection{Comparison with PHANGS-HST Pipeline Results}

\begin{figure}[h]
    \centering
    \includegraphics[scale = 0.4]{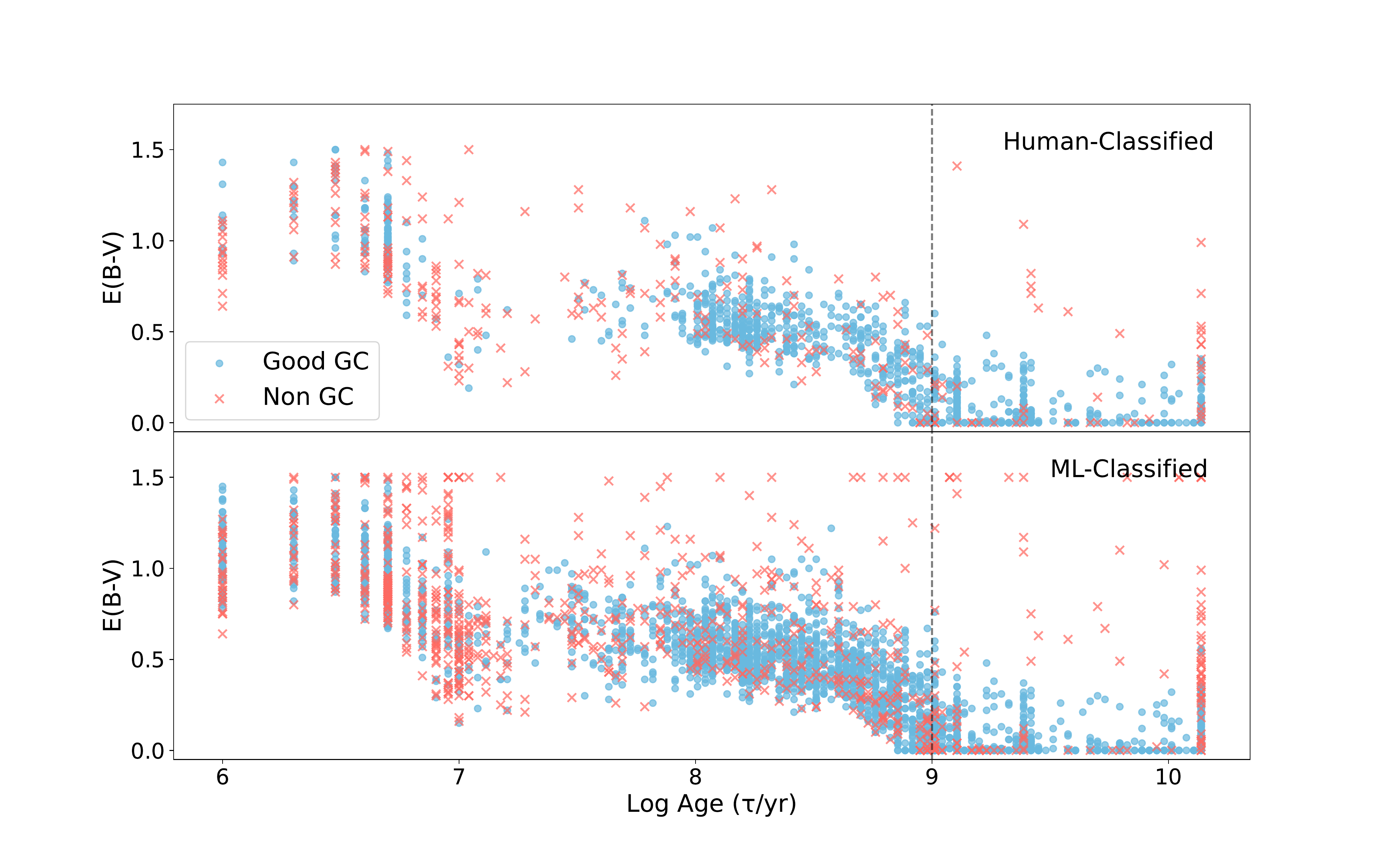}
    \caption{The best-fit age and reddening (E(B-V)) from the default PHANGS-HST pipeline for all color-selected GC candidates from the human-classified catalogs in the top panel and the ML-classified candidates in the bottom. Objects classified by us as likely GCs after visual inspection are shown in blue and contaminants are shown in red. We expect GCs to have estimated ages $\geq 1$~Gyr (dashed line), but only $\approx 25\%$  do.
    This illustrates the challenges of breaking the age-reddening-metallicity degeneracy using a single set of SSP model parameters, as discussed extensively in \citet{Whitmore23} and Thilker et al., in prep.} 
    \label{AGE_EBV}
\end{figure}

In Figure \ref{AGE_EBV} we plot the age vs reddening ($E(B-V)$) from the PHANGS-HST pipeline for all color-selected GC candidates. The top panel contains the human-classified candidates, with sources we believe are likely to be GCs after our visual inspection shown in blue and contaminants in red. In the figure, good (old) GCs can be found at almost any PHANGS-HST pipeline age, 
due to challenges in breaking the age-reddening-metallicity degeneracy when only broad-band colors are used in SED fitting (Whitmore et al 2023), and not because they are actually young. 
If we assume that ages of 1~Gyr and older are reasonable for GCs when solar metallicity is assumed, only $\sim 25\%$ of our likely GCs have a reasonable age. The default fitting procedure assumes solar metallicity and allows for reddening up to $E(B-V) = 1.5$~mag \citep{Turner21}.
Because many GCs in spirals are blue and metal-poor, they have bluer colors than predicted by the solar metallicity models, and 
end up being best fit by a young (intermediate) age with high (moderate) reddening.

\cite{Whitmore23} introduced a method to determine the ages of ancient GCs in spiral galaxies. This method relies on selecting likely GCs based on their color in a region similar to the 'old cluster clump' shown in Figure~2 here, and refitting for the age using a 1/50$\times$solar metallicity model with very low reddening E(B-V) $<0.1$~mag.  Using this method, \citet{Whitmore23} showed that most GCs were fit with ages $\geq 1 \: {\rm Gyr}$ (see also Thilker et al., in prep).

We compare sources in our final GC catalogs with 388 sources that were corrected for age using the method described above in 6 galaxies.
We consider a ``match" to be a source that we have classified as a likely GC that has a corrected age $\geq 1$~Gyr. We also consider it a match if we classified a source as a reddened young cluster and the corrected age is less than 1~Gyr. We find that 324 of the 388 matched sources have ages consistent with our classifications, or a $\sim$ 83$\%$ match.

\section{Discussion} \label{sec:discussion}
\par

\subsection{Discovery of A Faint Population of Globular Clusters in NGC~3627 and NGC~628}

The globular cluster luminosity function (GCLF) has been well studied in elliptical galaxies and has a universal shape with a peak near $M_V \approx -7.4$ mag and a Schechter-like downturn at the bright end. While the GCLF has not yet been well-studied in spiral galaxies, it appears to have a similar shape and peak luminosity in the Milky Way \citep[e.g.][]{Jordan07, Rejkuba12}, M31 \citep{Secker92,Reed94}, and M81 \citep{Cortez10, Lomeli-Nunez22}.  
The peak and faint end of the luminosity function are likely shaped by cluster disruption, since lower mass (luminosity) clusters are preferentially disrupted due to the evaporation of stars caused by two-body relaxation on Gyr timescales \citep[e.g.][]{Fall01}. 
The GC candidates in the spiral galaxy M101 present an exception to this universal shape \citep{Barmby06, Simanton15}. M101 has an excess of faint, red clusters which have all the expected photometric and morphological properties of GCs, but which have a luminosity function that continues to increase (rather than turnover) in a power law fashion down to at least one magnitude fainter than the expected peak, $M_V \approx -6.4$~mag.

\begin{figure}
    \centering
    \includegraphics[scale= 0.45]{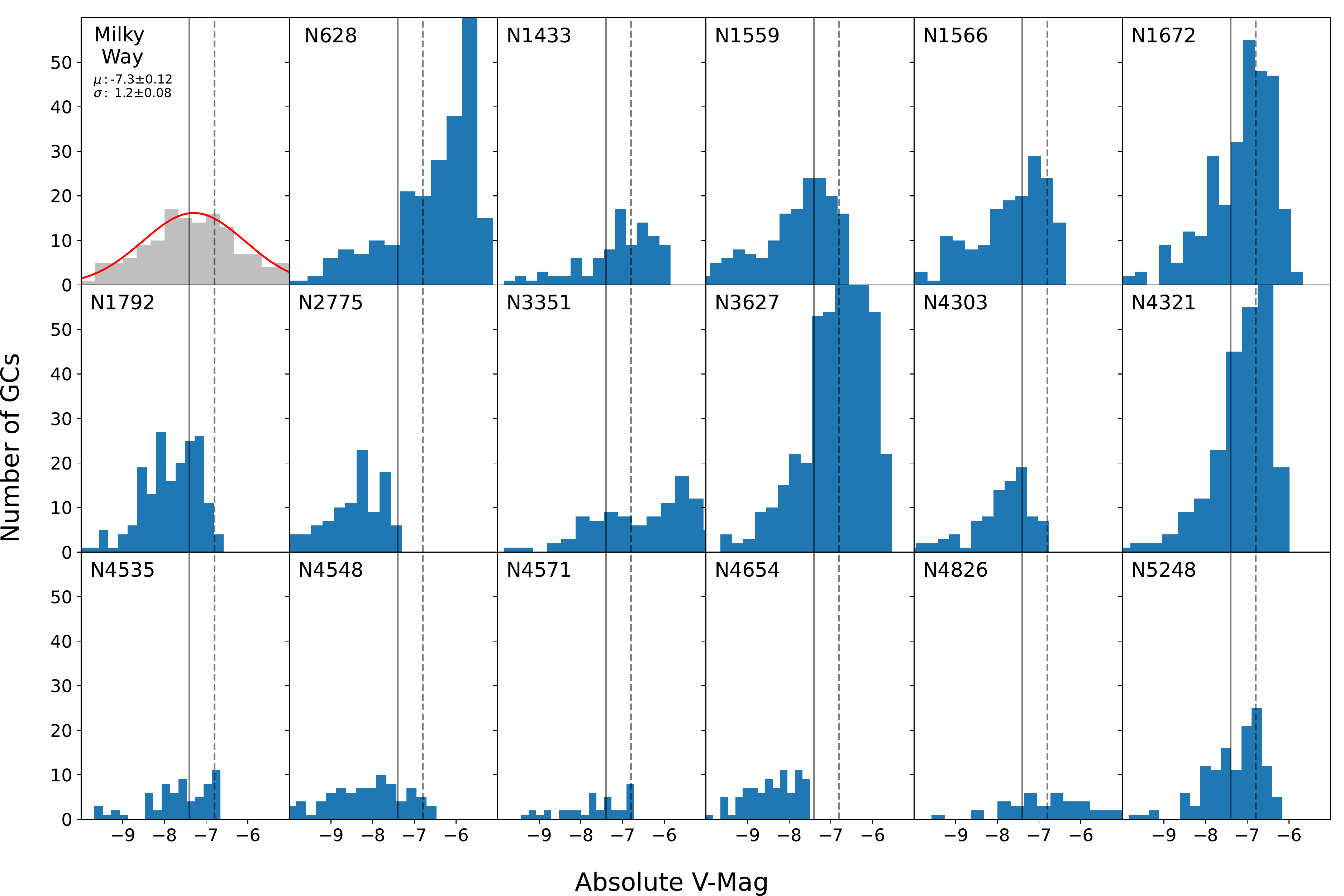}
    \caption{ Globular cluster luminosity functions are shown as absolute V-band magnitude histograms for each galaxy. The accepted universal turnover is marked with a solid vertical line at $M_V = -7.4$ mag, while a dashed vertical line is used to mark the magnitude limit of $M_V=-6.8$ that we adopt to divide between ''faint" and ''bright" cluster popluations. The top left panel shows the luminosity function of GCs in the Milky Way\citep{Harris96, Harris10}.  We fit a Gaussian to the luminosity function of GCs in the Milky Way, where the V band magnitudes are taken from the most recent version of the MW GC catalogs.  The best fit, shown in red in the top-left panel, has a peak near $M_V \approx -7.30\pm0.12$~mag and a dispersion parameter $\sigma=1.20\pm0.08$~mag, very similar to the parameters found earlier by \citet{Secker92} for the Milky Way GCLF. }
    \label{fig:Luminosity_func_counts}
\end{figure}

Figure~\ref{fig:Luminosity_func_counts} shows V-band luminosity histograms for all the GC systems studied in this work. We use bins of equal-width in magnitude in the range from $-$10 to $-$5.5 mag. The expected turnover $M_\mathrm{V} \approx -$7.4 mag is represented by the dashed black line in each panel. The upper left panel uses V-band magnitudes for Galactic GCs \citep{Harris96, Harris10}, and the best fit model Gaussian. The best fit gives a peak at $M_V = -7.30 \pm 0.12$~mag, and a standard deviation $\sigma = 1.20\pm0.08$, which is in good agreement with the earlier fit results for the Milky Way GC system from \citep{Secker92}, who found a
peak of $M_V=-7.29\pm13$~mag and dispersion $\sigma = 1.1\pm0.1$~mag.

The luminosity functions have not been corrected for completeness, although by including the machine learning-classified catalogs the distributions go well below the expected turnover in the majority of cases. In 10 galaxies the catalogs extend $\sim1$ magnitude fainter than the expected turnover, and in 5 of these cases, they extend a full 1.5 magnitudes beyond the expected turnover. There are seven galaxies where the cluster catalogs do not reach at least 1~magnitude below the expected turnover: NGC 1792, NGC 2775, NGC 4303, NGC 4535, NGC 4548, NGC 4571, and NGC 4654. These are some of the more distant galaxies in our sample, ranging from 15.8--23.15 Mpc. In addition, NGC\,4826 has poor statistics, making it difficult to assess the shape of its luminosity function. We will not consider these 8 galaxies further. 

The shapes of the GCLFs in the 9 galaxies which reach at least an $M_V$ of $-$6.4 fall into three categories: (1) we see a clear peak that is consistent with $M_V \approx -$7.4, (2) there may be a peak near the expected value, but the distribution is broad and the peak is indistinct, and (3) the number of clusters continue in power-law fashion well beyond the expected peak luminosity. 
We find 2 galaxies which fit into the first category: NGC 1559 and NGC 1566. Three galaxies fit into the second category: NGC 1433, NGC 3351, and NGC 5248, and 4 galaxies in the third category: NGC 628, NGC 1672, NGC 3627, NGC 4321. 
 
Galaxies in the third category where the GCLF continues to increase 
to at least 1 magnitude below the expected turnover luminosity are of great interest. M101 is the only galaxy to date known to convincingly show this distribution of GC luminosities. Here, we find that NGC~628 and NGC~3627 also have large, convincing populations of faint ($M_V \geq -6.8$) GCs, with 219 and 350 of these sources respectively.

\begin{figure}
    \centering
    
    \includegraphics[width = \textwidth]{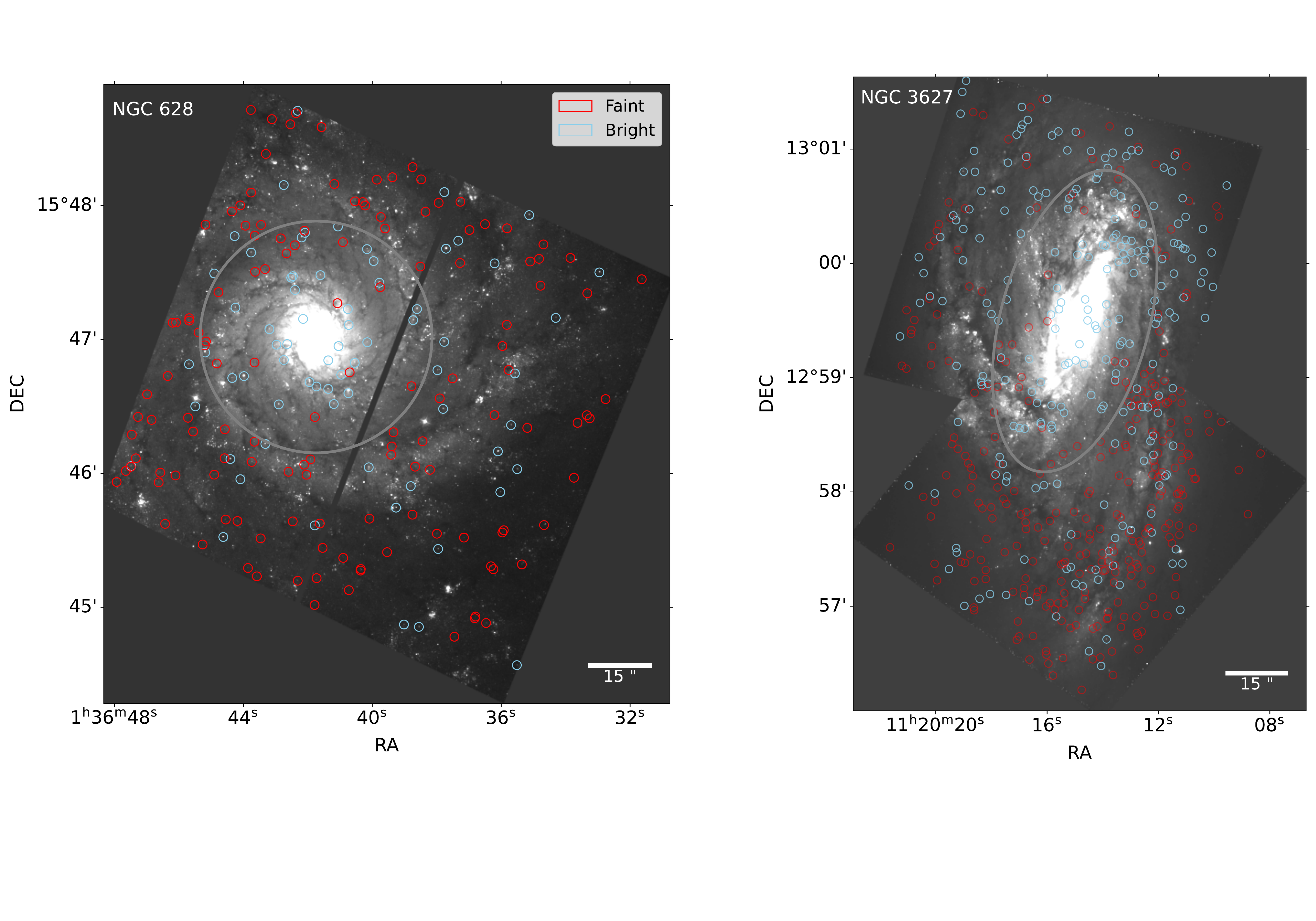}
    \vspace{-2cm}
    \caption{The locations of all likely globular clusters are  overplotted on a $V$-band image of their host galaxy, NGC 628 (left) and NGC 3627 (right). 
    The bright globular clusters ($M_V < -$6.8) are indicated by blue circles and faint ones ($M_V > -$6.8) are in red. The ellipses show the approximate distance where the radial density of the faint GC population peaks: 
    2.5~kpc for NGC 628 and 4.5~kpc for NGC 3627.}
    \label{fig:spatial_dist_all}
\end{figure}

\subsection{What is the Nature of the Faint Globular Clusters?}

In this section we focus on the large populations of faint GCs found in NGC~628 and NGC~3627 ($M_V \geq-$6.8). 

We consider three possibilities for the nature of these faint cluster populations: (1) they are young, reddened clusters in the disk, (2)  intermediate age clusters ($\sim$ a few 100~Myr to several Gyr) with low to moderate reddening also in the disk, or (3) ancient ($\geq$ 10~Gyr) clusters in either the halo or disk.  

We believe that the first possibility, namely that these faint GC candidates are actually very young, reddened clusters is quite unlikely, since none have associated H$\alpha$ emission. In NGC~628, we find that only 10 of the 124 (8\%) fainter than the magnitude limit of the human-selected catalogs even have H$\alpha$ emission within $\approx50$~pc, and in these cases, the emission appears to be associated with another nearby object. The lack of nearby H$\alpha$ emission for this population of faint red clusters indicates that they are unlikely to be very young, reddened clusters. 

\begin{figure}[h]
    \centering
    \includegraphics[scale = 0.5]{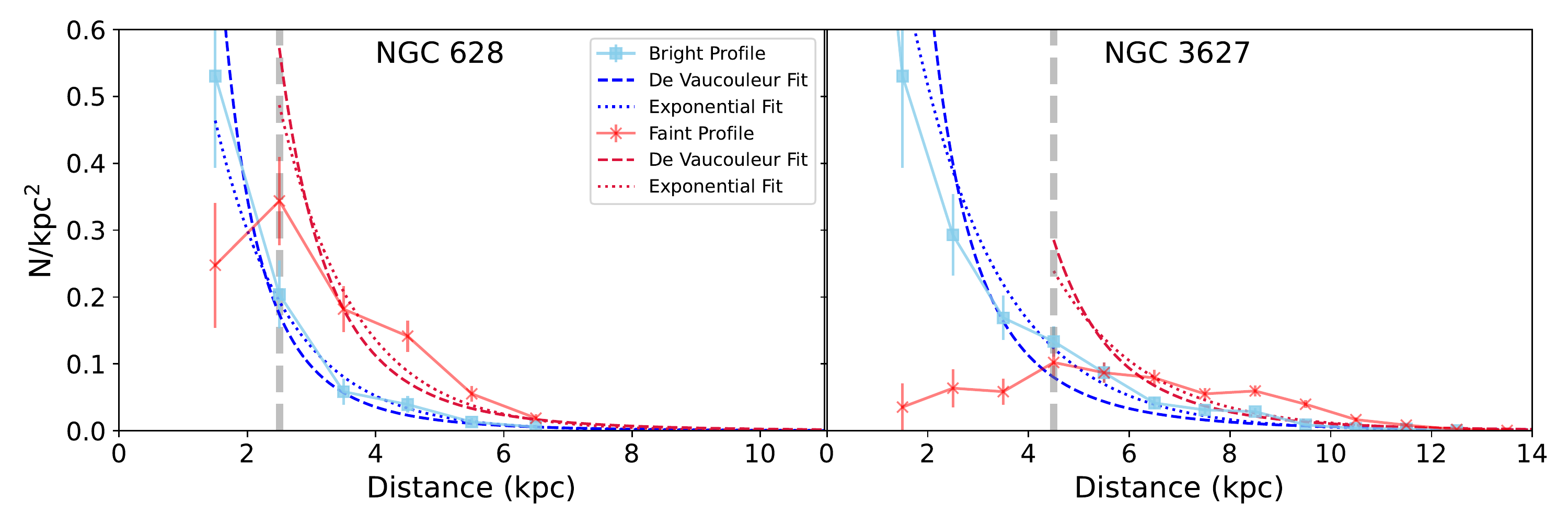}
    \caption{
    Radial density distributions of bright (blue) and faint (red) globular clusters (divided at $M_V=-6.8$) are plotted using equal-size radial bins.  The profiles for NGC~628 are on the left and for NGC~3627 on the right.   Error bars are calculated as Poissonian statistics. 
    The best fit de Vaucouluer (dashed) and exponential (dotted) profiles to each distribution are shown. 
    }
    \label{fig:Spat_dis}
\end{figure}

So, we focus on the remaining 2 possiblities by studying the spatial distribution of GCs, in order to discern where in the galaxy they likely reside. Figure \ref{fig:spatial_dist_all} shows the locations of faint (red, $m_V \geq -6.8$~mag) and bright (blue, $m_V < -6.8$~mag) GCs in NGC~628 and NGC~3627, while Figure~\ref{fig:Spat_dis} shows radial density profiles of each population. The error bars show Poisson statistics.
 
Higher background level in the bulge and central regions make it more difficult to identify faint GCs than bright ones.  The dashed vertical lines in Figure~12 are placed at the galactocentric distance where the radial profiles of the faint GCs peak; this occurs at 2.5~kpc in NGC~628 and 4.5~kpc in NGC~3627.  At closer distances, the profiles for the faint clusters decline, almost certainly due to incompleteness, while those for the bright ones continue to rise.  
At distances beyond the dashed lines in Figure~12, we see that the radial density profiles for the faint GCs are flatter and more extended than their bright counterparts.   
An exponential is typically used to describe spiral disks, and the de Vaucouleur profile is often used to describe a spheroidal population, like the central bulge \citep[e.g.,]{Baggett98}. Quantitatively, we fit the bright and faint distributions to both an exponential and a de Vaucouleur profile \citep{DeVaucouleurs}. The exponential is fit to the form: $log(y) = a_{\rm exp}x+b_{\rm exp}$ and the de Vaucouler profile is fit to the form: $ln(y) = a_{\rm Vauc}x^\frac{1}{4} + b_{\rm Vauc}$. 

The fits are shown in Figure \ref{fig:Spat_dis}, as indicated by the legend.  For the bright GC populations, there is little difference between the best de Vaucouleur and exponential fits (blue dashed and dotted lines, respectively), although the de Vaucouleur profile appears to better fit the inner-most data point in NGC~628.  Neither profile does a very good job of representing the faint GC populations (red curves).  While it is not possible to draw firm conclusions, the faint GCs clearly have a broader, more extended distribution than the bright GCs.
We speculate that the bright GCs are distributed in a spheroidal distribution, consistent with a galactic bulge$+$halo, while the faint populations may be associated with the disks of NGC~628 and NGC~3627.

\section{Conclusions}
\label{sec:Conc}

In this work, we have created new catalogs of globular clusters (GCs) in 17 spiral galaxies observed as part of the PHANGS-HST Treasury program.  
The galaxies have stellar masses between $\approx 10^{10}$ and $10^{11}~M_{\odot}$, star formation rates between 0.2 and 7.6~$M_\odot \mbox{yr}^{-1}$, include barred and unbarred Sa through Sd spirals. 
Only one of these galaxies (NGC\,628) has a previously published catalog of GCs. 
Initial candidate lists were selected by applying color cuts of B-V $\geq 0.5$~mag and V-I$\geq 0.73$~mag to two distinct cluster catalogs, one based on morphological classification by co-author B. Whitmore and the other by a machine-learning algorithm \citep{Wei20, Hannon23}.  The machine-learning based catalogs reach $\approx 0.5-2.0$~mag fainter  than the human ones, and the color selection is based on the intrinsic colors of globular clusters in the Milky Way.


Each color-selected candidate GC 
was subjected to a visual inspection which 
included morphology in B, V, and I filters, an assessment of whether or not it is associated with H$\alpha$/CO emission or dust, and the local environment to determine if the source is a GC or a contaminant (e.g.\ reddened young or intermediate age cluster, background galaxy, or foreground star). 
We expect GCs to be round (dynamically relaxed), not near regions of recent star formation, and uniform in color. 
Some key statistics related to the production of the GC catalogs are given below.

\begin{itemize}

\item We found that 2705 out of 3567 color-selected candidates from the PHANGS-HST catalogs are likely GCs. 

\item The percentage of confirmed GCs in the ''bright" color-selected samples is similar in both the human and machine learning-based cluster catalogs,
$\sim$ 74.3--78.3\%.

\item There are a total of 1444 likely GCs identified in the machine learning catalogs that are fainter than the magnitude limits of the human-based catalogs.  

\item These 'faint' catalogs had $\approx79.8$\% confirmation rate, similar to that for the brighter samples.



\end{itemize}

For each globular cluster, we present their position, photometry 5 filters: F275W ($\approx$NUV), F336W($\approx$U), F435W/F438W( $\approx$B), F555W($\approx$V), F814W($\approx$I) bands, and their concetration index measured in 1 and 3 pixels.
These are the largest and most homogeneously created catalogs of candidate globular clusters in spiral galaxies to date.
The final catalogs created in this work can be found at https://dx.doi.org/10.17909/jray-9798.

We studied the luminosity functions of the GC systems in the 17 spiral galaxies, and found that most are 
consistent with the expected shape, an approximately 
log-normal distribution with a peak or turnover near $M_V \approx -7.4$~mag.  Two galaxies however, have luminosity distributions which continue in power-law fashion at least 1~mag fainter than the expected turnover: 
NGC\,628 and NGC\,3627.
These atypical luminosity functions are similar to that found for likely GCs in M101, making NGC\,628 and NGC\,3627 only the second and third spiral galaxies known to have these faint excess populations of red, old clusters.

The faint GC candidates in NGC~628 and NGC~3627 have a more extended spatial distribution than their bright counterparts, as expected from a disk population.  They are unlikely to be reddened, young disk clusters because they do not show H$\alpha$ emission and are not within obvious dust lanes.  Their radial distributions are best fit by an exponential profile with a weak excess coincident with the location of spiral arms.  Clusters with ages $\approx \mbox{few} \times 100$~Myr have been found to still be associated with the inner spiral arms of M51.  We therefore speculate that the faint populations in NGC~628 and NGC~3627 could be lower-mass, intermediate-age ($\approx1$ to a few Gyr)
clusters that formed and survived in the disk.  

Follow-up, optical spectroscopy would better constrain the ages and dynamics of the faint GC candidates we have found in NGC\,628 and NGC\,3627.
In an upcoming work, we will use our new catalogs to establish the color distributions and calculate specific frequencies of the GC systems for the spiral galaxies studied here. 

\acknowledgments

We dedicate this paper to the memory of Julie Whitmore (January 13, 1954 - August 23, 2023).

This work is based on observations made with the NASA/ESA Hubble Space Telescope, obtained at the Space Telescope Science Institute, which is operated by the Association of Universities for Research in Astronomy, Inc., under NASA contract NAS 5-26555. These observations are associated with program \#13364. 

This work was carried out as part of the PHANGS collaboration.  R.C. acknowledges support from NSF grant 1517819. 

K.G. is supported by the Australian Research Council through the Discovery Early Career Researcher Award (DECRA) Fellowship (project number DE220100766) funded by the Australian Government. 
K.G. is supported by the Australian Research Council Centre of Excellence for All Sky Astrophysics in 3 Dimensions (ASTRO~3D), through project number CE170100013.
\software{\texttt{Photutils} \citep{photutils}, \texttt{Matplotlib} \citep{matplotlib}, \texttt{NumPy} \citep{numpy-guide,numpy}, \texttt{Astropy} \citep{astropy13,astropy18,astropy22}, \texttt{SciPy} \citep{scipy_new}, \texttt{SAOImage DS9} \citep{ds9}}
\bibliography{master}

\end{document}